\documentclass[aps,prl,reprint,longbibliography]{revtex4-1}
\usepackage{amsmath}
\usepackage[utf8]{inputenc}
\usepackage{amsmath,amssymb,mathtools}
\usepackage{graphicx}
\usepackage[hidelinks, pageanchor=false]{hyperref}
\usepackage{xcolor}
\usepackage{multirow}
\usepackage{textgreek}
\usepackage{babel}
\usepackage{xspace}
\usepackage{enumerate}
\usepackage{caption}
\usepackage{ragged2e}

\usepackage{algorithm}
\usepackage{algpseudocode}
\usepackage{relsize}

\usepackage{float} 
\usepackage{comment}
\usepackage{orcidlink}
\usepackage[
  a4paper,
  twoside=false,
  left=1.7cm,
  right=1.7cm,
  top=2.5cm,
  bottom=3cm
]{geometry}

\renewcommand{\thefigure}{\arabic{figure}}  
\allowdisplaybreaks

\def\beq{\begin{equation}}
\def\eeq{\end{equation}}
\def\bea{\begin{align}}
\def\eea{\end{align}}

\def\Eq#1{Eq.~(\ref{#1})}

\def\ra{\rangle}

\def\ket#1{|#1\ra}

\def\beq{\begin{equation}}
\def\eeq{\end{equation}}
\def\bea{\begin{eqnarray}}
\def\eea{\end{eqnarray}}
\def\beqn{\begin{eqnarray}} \def\eeqn{\end{eqnarray}}
\def\Eq#1{Eq.~(\ref{#1})}

\def\qon#1{q_{#1,0}^{(+)}}

\def\qb{\mathbf{q}}

\def\lb{\boldsymbol{\ell}}

\footnotetext{k.pyretzidis@csic.es}

\begin{document} 
%\begin{titlepage}
\author{Konstantinos Pyretzidis \orcidlink{0009-0002-0570-582X}$^{*, \dagger }$} 
\author{Jorge J. Mart\'{\i}nez de Lejarza\orcidlink{0000-0002-3866-3825}$^{*}$} 
\author{Germán Rodrigo\orcidlink{0000-0003-0451-0529}$^*$} 
\affiliation{%
 $*$Instituto de F\'{\i}sica Corpuscular, Universitat de Val\`encia - Consejo Superior de Investigaciones Cient\'{\i}ficas, Parc Cient\'{\i}fic, E-46980 Paterna, Valencia, Spain} 

\title{
Unlocking Multidimensional Integration with Quantum Adaptive Importance Sampling}

\date{\today}

\begin{abstract}

Multidimensional numerical integration is a central ingredient of theoretical predictions in high-energy physics, where multiloop Feynman diagrams and phase-space integrals are computationally demanding due to divergences and complex mathematical structures. Established Adaptive Importance Sampling methods for numerical integration, such as VEGAS, iteratively refine a grid in a separable way, dimension by dimension. This keeps the algorithm scalable but reduces performance when strong inter-variable correlations are present. In this work, we introduce a hybrid quantum-classical algorithm that performs Quantum Adaptive Importance Sampling (QAIS) for multidimensional Monte Carlo integration. Our approach uses a Parametrized Quantum Circuit to encode a non-separable Probability Density Function on a multidimensional grid and allocate samples efficiently in the integration domain. We apply the method to a sharply peaked loop Feynman integral and to multi-modal benchmark integrals. Our results show that QAIS provides an efficient route for high-precision evaluation of multidimensional integrals.

\end{abstract}

\preprint{\today}

\maketitle

% \section*{Summary Sentence (up to 350 characters)}

% Theoretical calculations in particle physics rely on multidimensional integrals that become difficult to evaluate in the presence of correlations. We show that a hybrid quantum-classical algorithm implementing Quantum Adaptive Importance Sampling can learn correlated structures efficiently and deliver precise results for nontrivial integrals.

%%%%%%%%%%%%%%%%%%%%%%%%%%%%%%%%%%%%%%%%%%%%%%%%%%%%%%
%%%%%%%%%%%%%%%%%%%%%%%%%%%%%%%%%%%%%%%%%%%%%%%%%%%%%%
\section*{Introduction}
% \label{sec:Introduction}
%%%%%%%%%%%%%%%%%%%%%%%%%%%%%%%%%%%%%%%%%%%%%%%%%%%%%%
%%%%%%%%%%%%%%%%%%%%%%%%%%%%%%%%%%%%%%%%%%%%%%%%%%%%%%

% \rd{This is ready for arxiv. I reposiitoned the plots, and included the Supplementary Material . I suppose I will wait for the DOI. If you want to make any change, pls let me know }

Perturbative Quantum Field Theory has long been established as the leading method for making precise theoretical predictions of observables at high-energy particle colliders. By systematically expanding in the interaction couplings, scattering amplitudes and differential cross sections at high perturbative orders involve loop and phase-space integrals of increasing dimensionality. As experimental facilities such as the CERN’s Large Hadron Collider~(LHC) achieve unprecedented measurement precision, there is a growing demand for equally accurate theoretical calculations to ensure a meaningful comparison between theory and experiment.

Therefore, in the context of particle physics phenomenology, there is an extensive need for more efficient theoretical frameworks and computational strategies. The primary computational bottleneck lies in the numerical integration of the high-dimensional functions that arise from the perturbative expansion. In particular, achieving the sub‐percent precision required by modern experiments is challenging, as the required number of integrand evaluations grows rapidly with dimensionality. This computational burden is exacerbated by the presence of integrable but numerically unstable singularities, sharp resonant peaks, and oscillatory regions, all of which require very fine resolution and large samples in localized subdomains, though these regions constitute only a tiny fraction of the total integration space.

These issues make brute-force Monte Carlo (MC) integration computationally prohibitive, and in some cases, entirely impractical, motivating the development of sophisticated variance reduction techniques such as Importance Sampling (IS). The most notable example is the VEGAS~\cite{Lepage:1977sw,Lepage:123074,Lepage:2020tgj} algorithm, which performs Adaptive Importance Sampling (AIS) with a hyper-rectangular grid, and has become the common standard. Specifically, VEGAS operates by iteratively adapting a piecewise-constant grid that concentrates samples in regions of high variance, dramatically reducing statistical uncertainties for a given budget of function evaluations. VEGAS and its variants are assimilated into mainstream libraries (e.g. Cuba~\cite{Hahn:2004fe}, MadGraph~\cite{Alwall:2011uj}, and Sherpa~\cite{Gleisberg:2008ta} among others). More recently, integrators based on Machine Learning (ML) have emerged \cite{Gao_2020,Gao_2020b,Bothmann:2020ywa,Winterhalder:2021ngy,Heimel_2023,Heimel_2024,Heimel_2025}, as well as VEGAS GPU-based optimizations to reduce runtime \cite{Carrazza:2020rdn}.  

As classical integrators approach their practical limits, attention has shifted towards Quantum Computing (QC), driven by theoretical advances in the field through quantum algorithms such as Shor’s integer factoring~\cite{Shor_1997}, Grover’s unstructured search~\cite{grover1996fastquantummechanicalalgorithm}, and the Harrow-Hassidim-Lloyd quantum linear solver~\cite{Harrow_2009}, as well as advancements in the quantum hardware. Quantum algorithms for accelerating MC methods have been developed recently~\cite{deLejarza:2023qxk,deLejarza:2024pgk,deLejarza:2024scm,Rodrigo:2024say,Agliardi:2022ghn,Williams:2025hza,Cruz-Martinez:2023vgs,akhalwaya2023modularenginequantummonte,Plekhanov_2022,Layden_2023,Mazzola_2021}. The majority of the integration approaches exploit the quadratic speedup offered by the fault-tolerant algorithm Quantum Amplitude Estimation~(QAE)~\cite{Brassard_2002,Montanaro_2015}.

In this paper, we leverage one of the key strengths of quantum computers, namely their ability to sample from non-trivial probability distributions. Examples include distributions that are hard to sample from classically, such as those generated by Instantaneous Quantum Polynomial (IQP) circuits \cite{Bremner_2010,Bremner_2016}. Notably, one of the few demonstrations of quantum advantage~\cite{Arute:2019zxq}, where a quantum device outperforms any classical counterpart, was fundamentally a sampling problem. This experiment highlighted the potential of quantum computers to surpass classical systems in specific tasks. With the current surge of generative modeling as a cornerstone of classical ML, and sampling being one of the strongest features of quantum devices, the intersection of these fields has attracted significant attention within the QC community~\cite{Romero2021,Paine2021,Zhu2022,Kyriienko2024,kasture2022protocols,rudolph2023trainability}. Notable examples include Quantum Generative Adversarial Networks~(QGANs)~\cite{Lloyd2018,Dallaire-Demers2018,Zoufal2019,Huang2021PRAppl}, Quantum Boltzmann Machines~(QBMs)~\cite{Zoufal2021,Coopmans2024}, and Quantum Circuit Born Machines~(QCBMs)~\cite{Liu2018,Benedetti2019npj,Benedetti_2020,Kiss2022,hibat2024framework}. Among other applications, quantum generative modeling has shown particular potential across a wide range of problems within High-Energy Physics (HEP)~\cite{Delgado:2022tpc,Delgado:2023ofr,Delgado:2024vne,Bermot:2023kvh, Tuysuz:2024hyl,Mart_nez_de_Lejarza_2025}, alongside other promising Quantum Machine Learning (QML) approaches that have also been explored~\cite{lejarza,deLejarza:2022vhe,Belis:2023atb,Schuhmacher:2023pro,Belis:2024guf, Belis:2021zqi,Casals:2025cuc}.

In this context, this paper aims to further expand the growing applications of QC in HEP. Specifically, we present a general-purpose hybrid (quantum-classical) algorithm for high-precision numerical integration of high-dimensional functions, that performs Quantum Adaptive Importance Sampling~(QAIS).  Working within the constraints of the Noisy Intermediate-Scale Quantum~(NISQ) era, we intentionally avoid fully fault-tolerant routines.  Instead, we focus on classical IS schemes motivated by HEP, which already explore exponentially large domains but suffer from the curse of dimensionality, as reflected in the scaling of uncertainties for certain integrals with increasing dimensionality. Our approach uses a Parametrised Quantum Circuit (PQC) to encode a non-uniform proposal Probability Density Function (PDF) on a grid. We benchmark the method against VEGAS, focusing on the accuracy and scalability of the integral estimate, for a sharply peaked loop Feynman integral and multi-peaked benchmark integrals. We evaluate QAIS under noiseless statevector simulations in order to isolate the algorithmic and statistical behavior of the method, serving as a proof of principle to assess the performance of the QAIS workflow. We find that quantum-generated proposal PDFs can provide accurate and robust estimates for challenging high-dimensional integration tasks.  A detailed study of finite shot training and hardware noise is left for future work, together with the implications of the induced stochasticity for the optimization  in a NISQ-oriented or early-fault tolerant implementation.

%%%%%%%%%%%%%%%%%%%%%%%%%%%%%%%%%%%%%%%%%%%%%%%%%%%%%%
%%%%%%%%%%%%%%%%%%%%%%%%%%%%%%%%%%%%%%%%%%%%%%%%%%%%%% 

%%%%%%%%%%%%%%%%%%%%%%%%%%%%%%
\begin{figure*}[t!]
    \includegraphics[width=\textwidth]{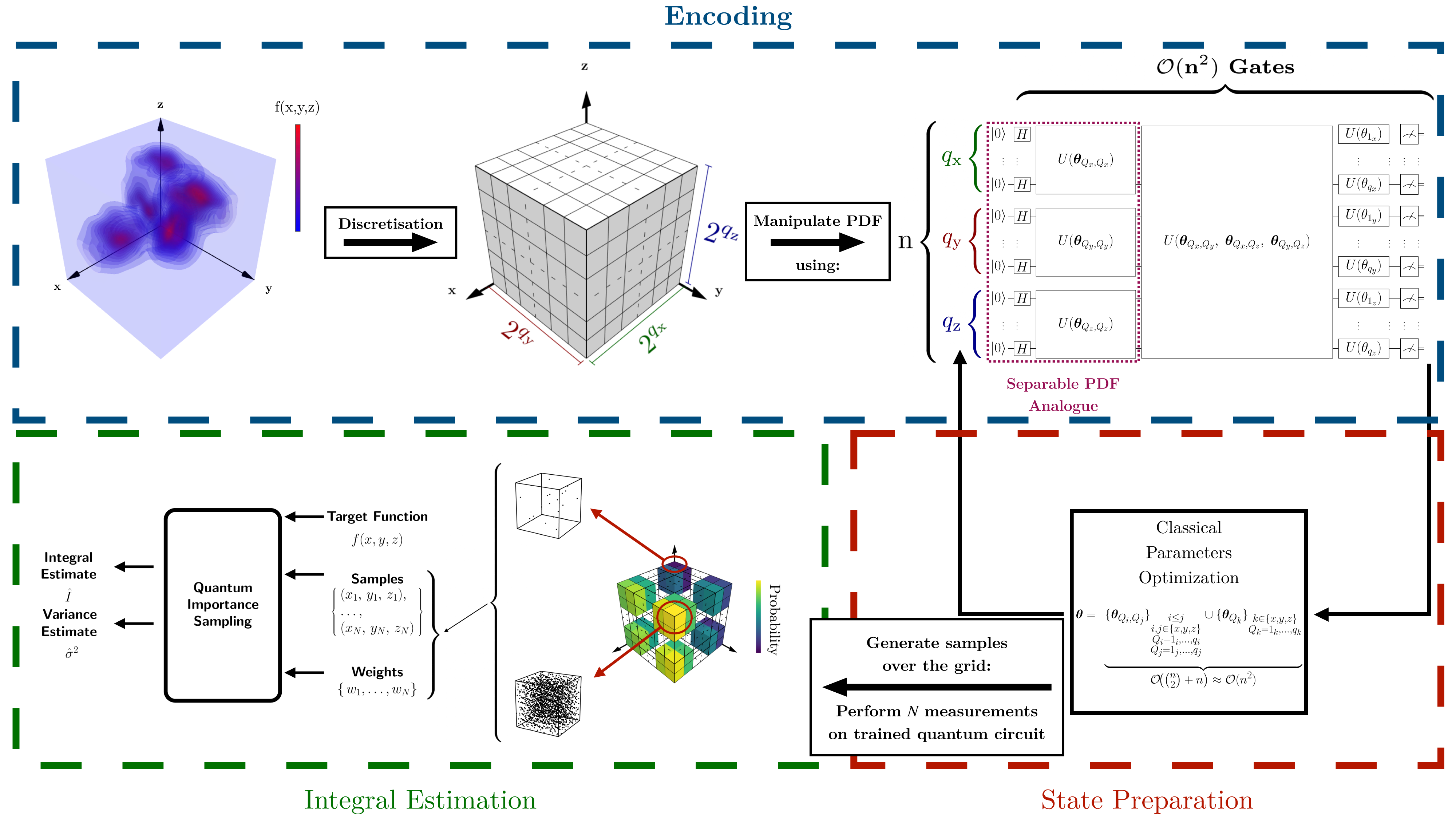}
    % need to change this
    \captionsetup{justification=Justified, singlelinecheck=off} 
    \caption{{ \bf Quantum Adaptive Importance Sampling workflow.} The Quantum Adaptive Importance Sampling algorithm consists of three main components: the Encoding, the State Preparation, and the Integral Estimation stages.} 
    \label{fig:summary}
\end{figure*}
%%%%%%%%%%%%%%%%%%%%%%%%%%%%%%

\section*{Results }

\subsection*{Algorithm Summary}

In this work, we introduce Quantum Adaptive Importance Sampling (QAIS), a hybrid quantum-classical algorithm that leverages a PQC to perform AIS for numerical integration of multidimensional functions. The central objective is to construct a PDF that accurately approximates the target integrand. In classical methods, such as VEGAS, the most computationally expensive step is evaluating the function at~$N$ distinct sample points. The aim of QAIS is to reduce the number of function evaluations needed to achieve the desired accuracy, by more efficiently allocating samples in the integration domain. It is crucial to note that this approach takes full advantage of the fact that MC integration demands minimal assumptions on the integrand. 

QAIS consists of three main elements, as presented in Fig.~\ref{fig:summary}. The Encoding stage, in which the PDF over the integration domain is discretized and mapped into a PQC. The State Preparation stage consists of adapting the parameters of a PQC to effectively shape the PDF generated by the quantum state to approximate the desired target function. In the third stage, the results from the $Z$-basis measurements of the optimal PQC are processed using a dedicated statistical framework.

The QAIS statistical framework is a modified version of conventional IS, adjusted to the quantum computational framework with the objective of fulfilling the requirements inherent to the efficiency of quantum measurements. In particular, finite-shot measurements sample only a subset of the computational basis states, and therefore only a subset of the discretized integration domain is supported by the proposal PDF. If this incomplete coverage is left untreated, it would introduce a systematic bias in the integral estimator. To address this issue, QAIS incorporates a debiasing strategy based on a Tiling algorithm, which reconstructs the non-Important Region support from the observed data while preserving the efficiency of the finite-shot measurement process. The full algorithmic and statistical framework and the details of the Tiling procedure are presented in Methods.

%%%%%%%%%%%%%%%%%%%%%%%%%%%%%%
\begin{figure*}[t!]
\begin{center}
\includegraphics[width=\textwidth]{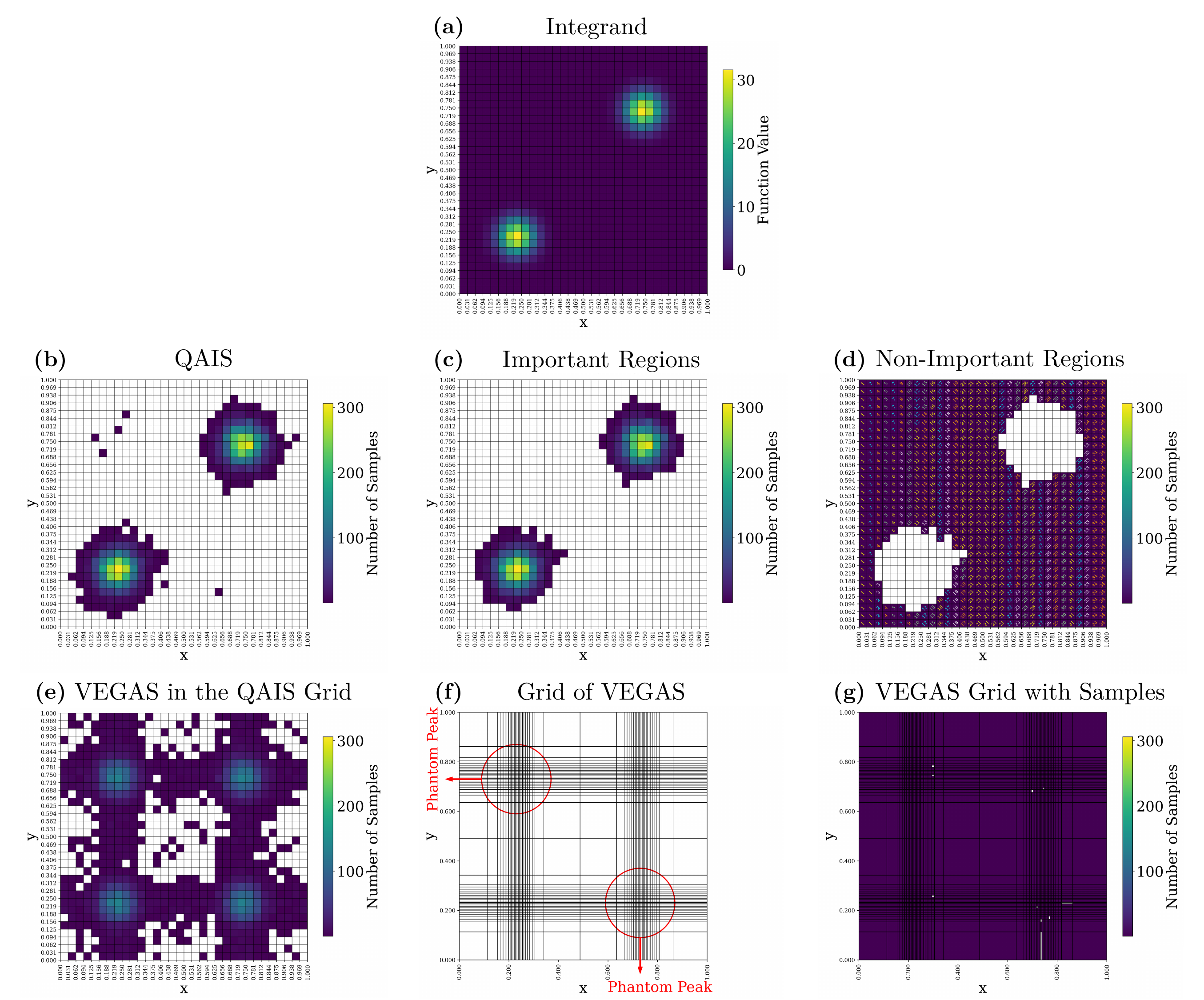}
\captionsetup{justification=Justified, singlelinecheck=off} 
\caption{ {\bf Sample allocation for a two-peak benchmark integrand.} {\bf (a)}~Integrand of~\Eq{eq:2D_gaussians_integral} on the discretized integration domain. {\bf(b)}~Samples obtained through Quantum Adaptive Importance Sampling (QAIS). {\bf(c)}~Regions classified as Important. {\bf(d)}~Non-Important Regions, subdivided into tiles. The number of each tile demonstrates its index, and the background color the number of samples. {\bf(e)}~VEGAS projected to the QAIS grid. {\bf(f)}~VEGAS on its native grid, with phantom peaks highlighted. {\bf(g)}~VEGAS on its native grid, with its corresponding number of samples per grid cell. For all cases, the number of samples used is $N=10^4$.
}
\label{fig:2D_comparison_plot}
\end{center}
\end{figure*}
%%%%%%%%%%%%%%%%%%%%%%%%%%%%%

%%%%%%%%%%%%%%%%%%%%%%%%%%%%%%
\begin{figure*}[t!]
\begin{center}
\includegraphics[width=\textwidth]{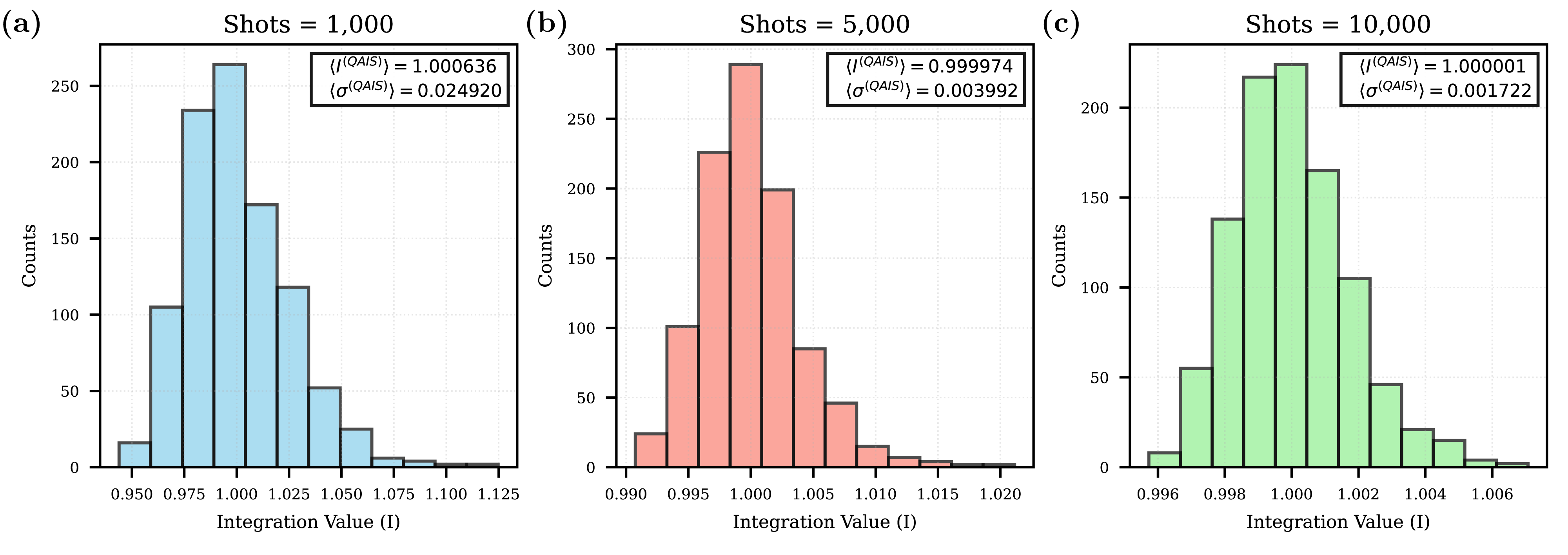}
\captionsetup{justification=Justified, singlelinecheck=off} 
\caption{{\bf Distribution of integration results as a function of the number of shots.} Integral estimations from $1000$ independent runs using the same proposal probability density function for the two-dimensional Gaussian integral in~\Eq{eq:2D_gaussians_integral}, for {\bf (a)} 1,000, {\bf (b)} 5,000, and {\bf (c)} 10,000 shots, respectively.
}
\label{fig:bias_check}
\end{center}
\end{figure*}
%%%%%%%%%%%%%%%%%%%%%%%%%%%%%

\subsection*{Absence of phantom peaks}

For a practical demonstration of the complete framework and the differences between QAIS and VEGAS, we present a two-dimensional toy integral composed of two Gaussian peaks in the diagonal. The integral is chosen for its simplicity, while still exhibiting nontrivial features for testing and illustrating the difference between the two methods. Specifically, we define:
\begin{equation}\label{eq:2D_gaussians_integral}
\int_{[0,0]}^{[1,1]} f(\mathbf{x}) d \mathbf{x} = \int_{[0,0]}^{[1,1]} \left( \sum_{i=0}^{1} e^{\left(-200 |\mathbf{x} - \mathbf{r}_i |^2\right)} \right) d \mathbf{x}~,
\end{equation}
where $\mathbf{r}_0 = (0.23, 0.23)$ and $\mathbf{r}_1 = (0.74, 0.74)$. Because the integral can be computed in closed form, it provides a clear benchmark for validating numerical methods. A normalization constant is chosen via analytical integration, so that the integral evaluates to $1$.

For QAIS, we have successfully trained a 10-qubit PQC to approximate the shape of $f(\mathbf{x})$ in~\Eq{eq:2D_gaussians_integral}. Figure~\ref{fig:2D_comparison_plot} clearly illustrates the distinction between Important and non-Important Regions, as described in the Methods subsection Debiasing Strategy using a Tiling Algorithm, since the target function contains large flat regions that contribute minimally, and it highlights differences of the two approaches. VEGAS relies on a strategy whose adaptation mechanism is based on manipulating grid boundaries, and does so separately for each dimension. In contrast, QAIS utilizes a stable grid, with equally spaced grid cells within each dimension.
The most crucial difference between the two methods is that QAIS encodes the number of samples directly into the amplitude of the PQC, providing a natural mechanism for sampling, while VEGAS employs a uniform PDF over an adapted grid. Then, its grid management routine effectively shifts samples around rather than sampling from complex high-dimensional distributions.

Furthermore, the PQC demonstrates the capability to accurately capture the function's details without throwing excessive samples in less relevant regions. Additionally, the Tiling algorithm is applied to separate the Important and non-Important Regions, generating an approximately monochromatic grid, which only appears in the non-Important areas.
The main advantage of QAIS, as illustrated in Fig.~\ref{fig:2D_comparison_plot}, is the absence of phantom peaks. In VEGAS, this is a consequence of separability resulting in a significant number of samples being wasted in irrelevant regions and considerably reducing the performance of the algorithm. 
It is relevant to mention that the smallest cell size produced by VEGAS in Fig.~\ref{fig:2D_comparison_plot} corresponds approximately to a discretization using seven qubits per dimension. Consequently, the VEGAS resolution is easily manageable with a PQC. 
 
To demonstrate that QAIS is unbiased, we conducted a series of nested MC experiments. Because the integral is analytically normalized to $1$, this exact value serves as the ground truth for validating our numerical estimates. For each experiment, we performed $1000$ independent integration runs, each with a fixed proposal PDF from the same trained PQC. We then repeated these experiments with different shot counts: $1000$, $5000$ and $10000$ samples. The results are shown in Fig.~\ref{fig:bias_check}. We see that the average integral value oscillates around the true value, improving accuracy progressively, and does not exhibit a systematic preference toward over or underestimation when the number of samples is varied. Additionally, the average QAIS standard deviation is presented. It shall be noted that estimates that deviate significantly from the mean integral value, always exhibit inflated standard deviation. Thus, QAIS produced accurate and unbiased estimates without the need to sample all $2^n$ possible states.

\begin{figure*}[t!]
\begin{center}
\includegraphics[width=\textwidth]{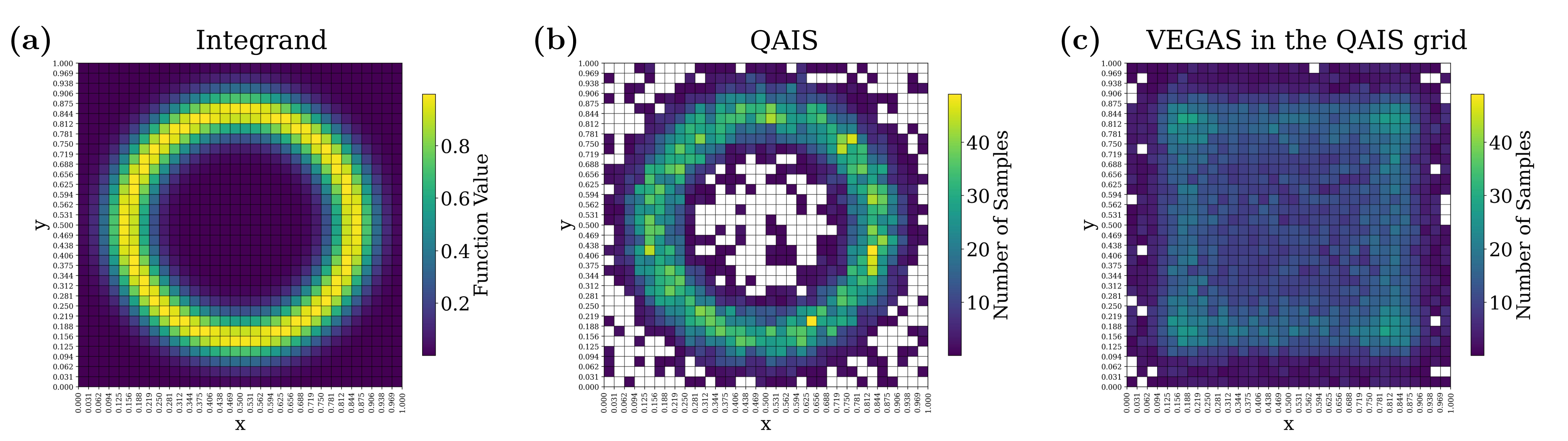}
\captionsetup{justification=Justified, singlelinecheck=off} 
\caption{ {\bf Ring-shaped benchmark integrand and sampling.} {\bf (a)} Integrand of \Eq{eq:ring_integral} on the discretized integration domain. {\bf (b)} Sample allocation by Quantum Adaptive Importance Sampling. {\bf (c)} Sample allocation by VEGAS, projected onto the Quantum Adaptive Importance Sampling grid. The total number of samples is $10^4$.}
\label{fig:ring}
\end{center}
\end{figure*}

Finally, we present a two-dimensional test case whose ring-shaped structure lies entirely outside VEGAS’s adaptive reach. The integral in closed form is given by
\begin{equation}\label{eq:ring_integral}
\int_{[0,0]}^{[1,1]} f(\mathbf{x}) d\mathbf{x}
=\int_{[0,0]}^{[1,1]} e^{ - 200 \bigl(| \mathbf{x}-\mathbf{r} | - 0.35\bigr)^2 }d\mathbf{x}~,
\end{equation}
where $\mathbf{r} = (0.5,0.5)$. Figure~\ref{fig:ring} illustrates the results for both methods. The exponential factor puts almost the entire weight of the integrand on a thin ring of radius $0.35$ centered at~$\mathbf{r}$, while in the rest of the integration domain, it is exponentially suppressed and effectively negligible. 

When VEGAS adapts the grid using the separable proposal of~\Eq{eq:projections}, it projects this PDF onto the coordinate axes, where the axis-wise viewpoint compresses the ring into two wide stripes, one in $x$ and one in $y$. As a result, VEGAS distributes its samples nearly uniformly across almost all of the domain. This leads to an overwhelming number of samples landing in the empty interior and exterior of the ring, with the highest local concentrations appearing in the four corners of the square, where the integrand is negligible. On the contrary, QAIS captures in detail the shape of the target distribution, and concentrates the samples precisely where the integrand is significant. This experiment illustrates how the projection based adaptation can fail when the integrand’s essential structure lies in correlations between integration variables, which motivates even further our approach of sampling from quantum states, which can naturally and efficiently encode such correlations through entanglement between registers.

%%%%%%%%%%%%%%%%%%%%%%%%%%%%%%
%%%%%%%%%%%%%%%%%%%%%%%%%%%%%

%%%%%%%%%%%%%%%%%%%%%%%%%%%%%%%%%%%%%%%%%%%%%%%%%%%%%%
%%%%%%%%%%%%%%%%%%%%%%%%%%%%%%%%%%%%%%%%%%%%%%%%%%%%%%
\subsection*{Application of QAIS to Multidimensional Integrals}
\label{sec:sec4}
%%%%%%%%%%%%%%%%%%%%%%%%%%%%%%%%%%%%%%%%%%%%%%%%%%%%%%
%%%%%%%%%%%%%%%%%%%%%%%%%%%%%%%%%%%%%%%%

We apply  QAIS on a set of illustrating multidimensional examples. First, we consider a one-loop Feynman integral with a single, very sharp peak. This example highlights the strengths of AIS because the narrow peak must be adequately located and precisely sampled, to achieve a reliable and accurate estimate with a reasonable number of function evaluations. In addition, this is an illustrating example to quantify how many computational basis states, out of the total exponential number of possible states, we must collect to capture with sufficient detail the most singular structures of the target function. We explicitly probe the same integrand for different number of qubits used in the discretization. In this context, we expect to get reliable and robust integral estimates within a reasonable amount of shots, and without gaining significant computational overhead due to the classical post-processing coming from the number of different basis states measured.

Next, we apply QAIS to integrals containing multiple sharp localized peaks, where we expect the non-separable proposal's strengths to appear. We consider test cases in which the underlying integrand structure consistently features the same set of peaks, although in a different number of dimensions. This setup lets us examine how increasing the dimension of the integration space affects the quality of the learned proposal PDF, and thus its ability to capture the integrand's correlations, and how the resulting integration uncertainty scales. 

Finally, we discuss how to extract meaningful insights from comparisons with VEGAS. Based on \Eq{eq:finalvegasestimateformulas}, VEGAS forms its final estimate, by combining information gathered from every proposal grid across all iterations, while assigning a greater weight to more accurate iterations. A full-scale direct comparison would require from our method a more efficient optimization procedure. However, as discussed briefly in the Methods subsection \hyperref[sec:sec3.4]{Structure and Optimization of the Parametrized Quantum Circuit}, optimization, especially for general purpose Ans\"atze, is still an open problem in QML. Thus, even though we have trained our PQCs and obtained high quality proposal PDFs, further studies are necessary for achieving an optimization procedure that is competitive with VEGAS. Therefore, we treat the trained PQC as a state preparation, where the PDF of the highly entangled quantum state provides a very accurate sampling allocation within the exponential sized space. We then compare the optimized QAIS proposal PDF with the best VEGAS grid. Thus, our main goal is to investigate whether a stable-sized PQC captures properly the intricate structures of the integrand and remains effective as the integration space grows. Additionally, we also aim to test whether and under which circumstances, the non-separable proposal PDF outperforms VEGAS's separable one, quantify the difference, and determine if it either reaches a target accuracy with fewer samples or achieves a more precise estimate within a fixed sample budget.

%%%%%%%%%%%%%%%%%%%%%%%%%%%%%%%%%%%%%%%%%%%%%%%%%%%%%%%%%%%%%%%%%%%%%%%%%%%%
\subsection*{One-loop pentagon Feynman integral in the Loop-Tree Duality}\label{sec:4.2.1}

%%%%%%%%%%%%%%%%%%%%%%%%%%%%%%
\begin{figure}[t]
\begin{center} 
\includegraphics[width=\columnwidth]{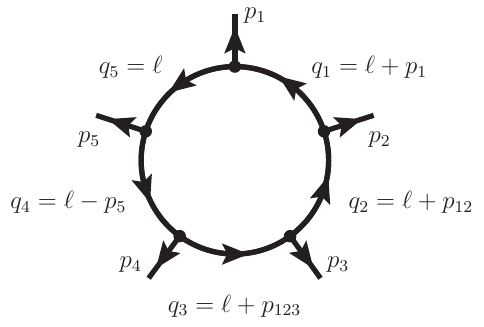}
\caption{\justifying{\bf Pentagon Feynman diagram at one loop.} The diagram shows the scalar five-point one-loop topology considered in this work, with loop momentum $\ell$, external momenta $p_i$, and internal propagator momenta $q_i$, where $p_{12}=p_1+p_2$ and $p_{123}=p_1+p_2+p_3$. Momentum conservation implies $\sum_{i=1}^5 p_i = 0$, considering all the external momenta outgoing.}
\label{fig:pentagon}
\end{center}
\end{figure}
%%%%%%%%%%%%%%%%%%%%%%%%%%%%%

%%%%%%%%%%%%%%%%%%%%%%%%%%%%%%
\begin{figure*}[t!]
\begin{center} 
\captionsetup{justification=Justified, singlelinecheck=off} 
\includegraphics[width=\textwidth]{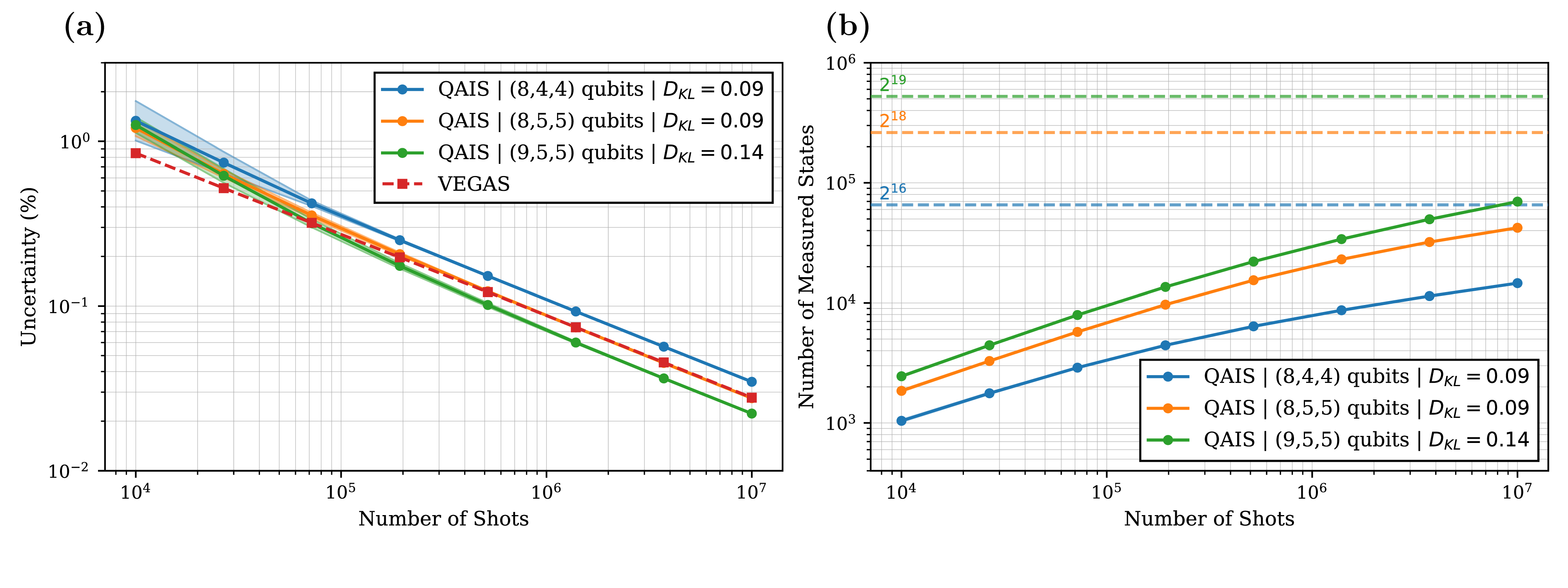}
\caption{{\bf One-loop pentagon Feynman integral.} {\bf (a) } Average uncertainty for 100 independent integration runs using the same proposal Probability Density Function (PDF) for the one-loop pentagon Feynman integral of Fig.~\ref{fig:pentagon}, across the full standard Monte Carlo sampling range. Each curve corresponds to a different discretization (16-19 qubits), with a proposal PDF optimized for that encoding. The shaded areas, correspond to the $\pm 1 \sigma$ dispersion of these 100 runs. The VEGAS result presented corresponds to its best iteration. {\bf (b) } Average number of distinct computational basis states measured in the same runs, as a function of the number of shots. }
\label{fig:pentagon_results0}
\end{center}
\end{figure*}
%%%%%%%%%%%%%%%%%%%%%%%%%%%%%

We consider the following one-loop scalar pentagon integral (see Fig. \ref{fig:pentagon}),
\beq
{\cal A}^{(1)}_5(\{p_i,m_i\}_{i=1}^5) = 
\int_\ell \prod_{i=1}^5 G_{\rm F}(q_i)~,
\label{eq:pentagon}
\eeq
where $G_{\rm F}(q_i) = (q_i^2-m_i^2+\imath 0)^{-1}$ are Feynman propagators  with mass $m_i$ and four-momenta 
$q_i = \ell + \sum_{j=1}^i p_j$,
with $q_5 = \ell$ by momentum conservation. The integration measure is $\int_\ell = -\imath \int d^4\ell/(2\pi)^4$, where $\ell$ is the loop four-momentum. The external four-momenta are $p_j$.

For the numerical implementation, we use the Loop-Tree Duality (LTD)~\cite{Catani:2008xa, Bierenbaum:2010cy,Bierenbaum:2012th,Buchta:2014dfa,Driencourt-Mangin:2019aix,Runkel:2019yrs,Aguilera-Verdugo:2019kbz,Runkel:2019zbm,Capatti:2019ypt,Capatti:2019edf,Bobadilla:2021pvr}. This representation is advantageous because the dimension of the integration domain is independent of the number of external particles, specifically, it is the three-dimensional Euclidean space of the loop three-momenta. The LTD representation is obtained by analytically integrating the energy component of the loop momentum by applying the Cauchy Residue Theorem. The resulting expression is also convenient because it is manifestly causal~\cite{Aguilera-Verdugo:2020set,Ramirez-Uribe:2024rjg,LTD:2024yrb,Imaz:2025buf,Rios-Sanchez:2024xtv,Aguilera-Verdugo:2020kzc,Ramirez-Uribe:2020hes,TorresBobadilla:2021ivx,Sborlini:2021owe,Kromin:2022txz,Ramirez-Uribe:2021ubp,Clemente:2022nll,Ramirez-Uribe:2024wua}, and so noncausal singularities of the integrand that lead to numerical instabilities are absent. In particular, the LTD representation  of the one-loop scalar pentagon in \Eq{eq:pentagon}, is given by
\beq
{\cal A}^{(1)}_5(\{p_i,m_i\}_{i=1}^5) = \int_{\lb}  \frac{1}{x_5} \sum \left( L_{ij}^+ L_{kl}^- + L_{ij}^- L_{kl}^+\right)~
, 
\label{eq:pentagonltd}
\eeq
 where $\int_{\lb} = \int_{\lb} \frac{d^3\lb}{(2\pi)^3}$, $x_5 = \prod_{i=1}^5 2 \qon{i}$ with $\qon{i}=\sqrt{\qb_i^2+m_i^2-\imath 0}$ the on-shell energies of the internal particles, and 
\beq
L_{ij}^\pm = \left( \frac{1}{\lambda^\pm_{i}} +
\frac{1}{\lambda^\pm_{j}} \right) \frac{1}{\lambda^\pm_{ij}}~,
\eeq
where the causal denominators are defined as 
\beq
\begin{split}
\lambda^{\pm}_{i} &=\qon{i}+\qon{i+1} \pm p_{i,0}~,  
\\
\lambda^{\pm}_{ij} &= \qon{i}+\qon{j+1} \pm \left(p_{i} + p_{j} \right)_0~, \qquad j = i+1~, \\ 
\lambda^{\pm}_{ij} &= \lambda^{\pm}_{i} + \lambda^{\pm}_{j}~,  \qquad \qquad \qquad \qquad \quad j = i+2~.
\end{split}
\eeq
It is understood that the indices of the on-shell energies are defined cyclically, namely $i=n+1$ with $n=5$ is equivalent to $i=1$. The integrand in~\Eq{eq:pentagonltd} depends on three independent integration variables. The loop momentum is parametrized in terms of the polar and azimuthal angles, $\lb = |\lb| (\sin \theta \cos \phi, \sin \theta \sin \phi, \cos \theta)$, and the modulus of the loop three-momenta is mapped to a finite interval, $|\lb| = z/(1-z)$ with $z\in [0,1]$

We consider the specific kinematic configuration P11 defined in~\cite{Buchta:2015wna} and compare our results with the numerical values obtained therein. This specific configuration corresponds to:
\beq
\begin{split}
p_1 &= (33.74515, 45.72730, 31.15254, -7.47943)~,\\
p_2 &= (31.36435, -41.50734, 46.47897, 2.04203)~,\\
p_3 &= (4.59005, 17.07010, 32.65403, 41.93628)~,\\
p_4 &= (29.51054, -28.25963, 46.17333, -35.08918)~,\\
m_1 &= m_2 = m_3 = m_4 = m_5 = 5.01213~.
\label{eq:p11buchta}
\end{split}
\eeq

\begin{figure*}[t!]
\begin{center} 
\captionsetup{justification=Justified, singlelinecheck=off} 
\includegraphics[width=\textwidth]{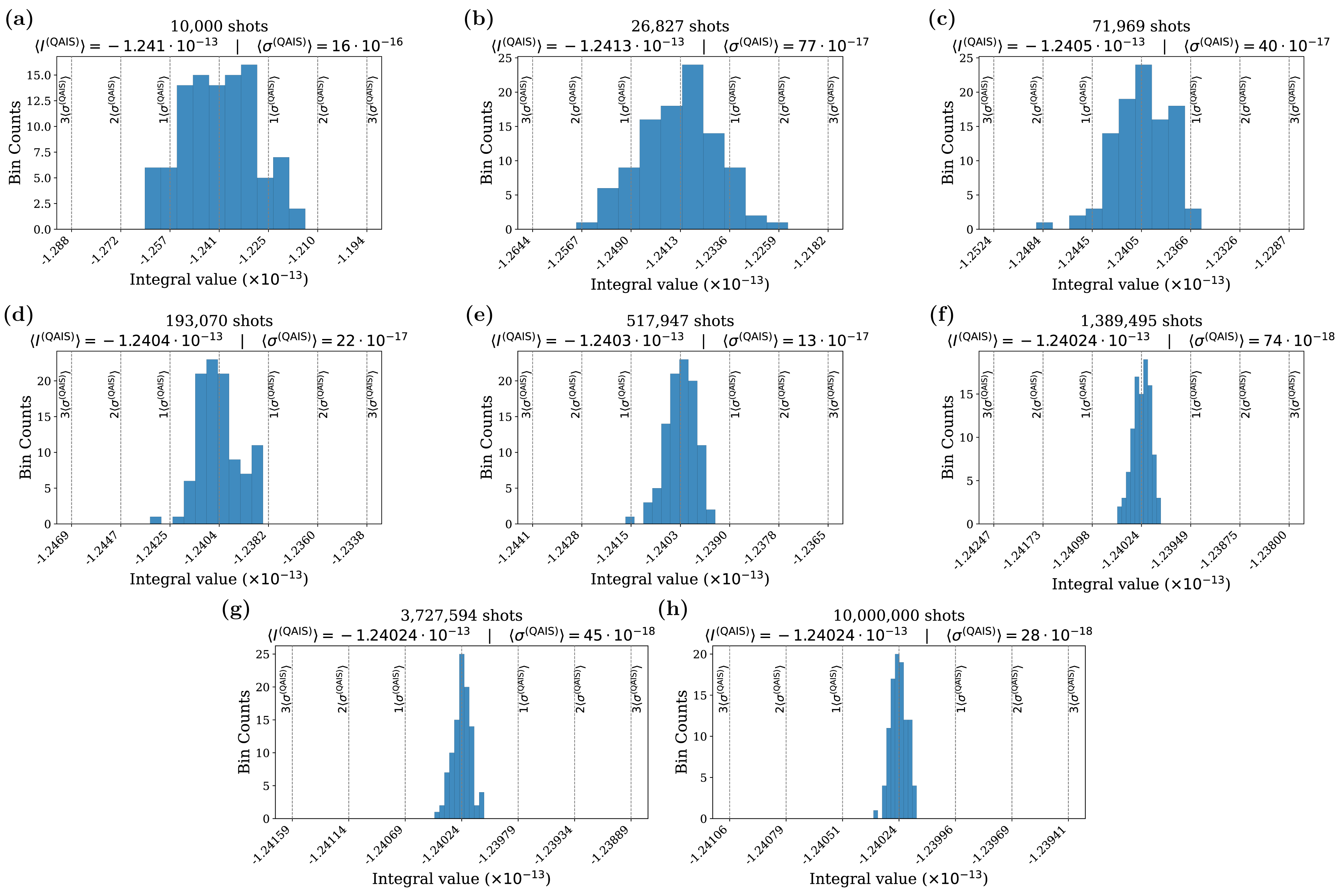}
\caption{{\bf Integral values for the one-loop pentagon Feynman diagram at (9,5,5) qubits as a function of the number of shots.}  {\bf (a)-(h)} histograms show the distribution of results corresponding to 100 independent integration runs with the parameter configuration in \Eq{eq:p11buchta}. The dashed vertical lines correspond to $1 \sigma$, $2\sigma$, and $3\sigma$ standard deviations. The reference value is obtained with VEGAS~\cite{Buchta:2015wna}, that is $I=-1.24027(16)\cdot 10^{-13}$.
}
\label{fig:pentagon_results}
\end{center}
\end{figure*}

The QAIS results are presented in Fig.~\ref{fig:pentagon_results0} and Fig.~\ref{fig:pentagon_results}. In Fig.~\ref{fig:pentagon_results0}~{\bf (a)}, we analyze the uncertainty of the integral estimate, defined as the ratio of the standard deviation divided by the integral estimate ($\sigma / I$), as a function of the number of shots. Across all shot counts, QAIS remains competitive with VEGAS, especially with increased discretization resolutions, although the overall performance difference of the two approaches is small across the entire shot count domain.

Figure~\ref{fig:pentagon_results0} {\bf (b)} shows the amount of quantum states observed from $Z$-basis measurements, out of the full Hilbert space, as a function of the number of shots. We observe a controlled gradual increase as the shot count rises. Since our integrands are typically nonzero everywhere and training is inherently imperfect, the number of contributing states naturally approaches the full Hilbert space dimension $2^n$ asymptotically. 

Nevertheless, the measurements populate only a small percentage of the Hilbert space, both in the sub-percent uncertainty region and at large shot counts. High accuracy therefore arises from sampling only a modest fraction of the available basis states. As the discretization is refined, that sampled fraction becomes even smaller relative to the enlarged Hilbert space, yet the quality of the estimate remains stable and improves slightly, underscoring the efficiency of direct sampling. Additional shots mainly revisit high impact cells that have already been identified, rather than diffusing across the entire space, showing that the trained PQC focuses on refining the IS weights where the integrand is most relevant and can achieve precision without exhaustive Hilbert-space coverage.

%%%%%%%%%%%%%%%%%%%%%%%%%%%%%%
%%%%%%%%%%%%%%%%%%%%%%%%%%%%%

Finally, the full integration results, run by run are presented in Fig.~\ref{fig:pentagon_results}. As the sample size increases the individual estimates cluster more tightly and do not exceed  one standard deviation, suggesting that our variance estimator is conservative.  This behavior is expected because quasi-random sequences achieve a Root Mean Square Error that asymptotically decays faster than the $N^{-1/2}$ rate of pseudo-random MC. For quasi-random sequences, in the worst case the error is $\mathcal{O}( N^{-1} (\log{N})^d )$ \cite{doi:10.1137/1.9781611970081}. However, because VEGAS, whose variance is given by \Eq{eq:variance_vegas}, serves as our reference, we report and use the conservative variance defined in~\Eq{eq:variance_qais}.

%%%%%%%%%%%%%%%%%%%%%%%%%%%%%%%%%%%%%%%%%%%%%%%%%%%%%%%%%%%%%%%%%%%%%%%%%%%%%%%%%%%%%%
\subsection*{Multi-peak Benchmark Integrals}\label{sec:4.2.2}

%%%%%%%%%%%%%%%%%%%%%%%%%%%%%%
\begin{figure*}[t]
\begin{center}
\includegraphics[width=.8\linewidth]{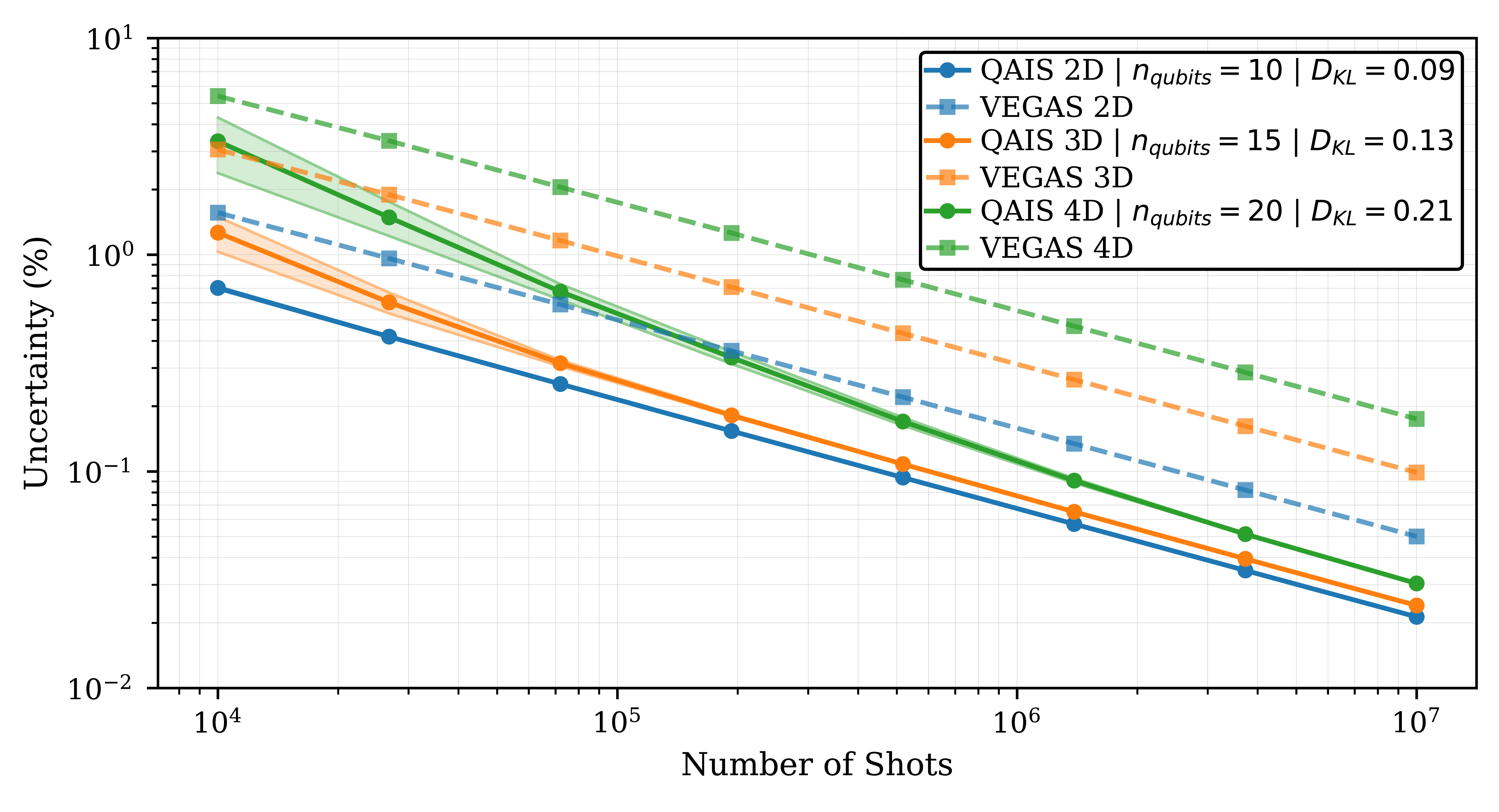}
\caption{\justifying {\bf Cross-dimensional integration comparison.} Average uncertainty for 100 independent integration runs using the same proposal Probability Density Function for the benchmark integral in~\Eq{eq:benchmark_int}, across the full standard Monte Carlo sampling range. The shaded areas, correspond to the $\pm 1 \sigma$ dispersion of these 100 runs. The VEGAS result presented corresponds to its best iteration.  }
\label{fig:4Dminimal}
\end{center}
\end{figure*}
%%%%%%%%%%%%%%%%%%%%%%%%%%%%%

Finally, we turn our attention to integrands with multiple peaks. This is a clear case where we expect the quantum‑generated PDF to surpass VEGAS's optimal grid in a considerable and scalable way, given that the phantom peaks problem intensifies as the number of integration dimensions rises. In order to demonstrate QAIS, we have chosen the benchmark integral from the VEGAS+ study~\cite{Lepage:2020tgj}. This benchmark integral contains three peaks along the diagonal. Because the projected shadows of the peaks do not cover one another in any dimension, it offers the most illustrative example of the phantom peaks problem. The analytic formula is:
\begin{equation}\label{eq:benchmark_int}
\int_{[0,1]^d} f(\mathbf{x}) d \mathbf{x} = \int_{[0,1]^d} \left( \sum_{i=0}^{2} e^{-50 |\mathbf{x} - \mathbf{r}_i |} \right) d \mathbf{x}
\end{equation}
where $\mathbf{r_0}=(0.23,\dots,0.23)$, $\mathbf{r_1}=(0.39,\dots,0.39)$, and $\mathbf{r_2}=(0.74,\dots,0.74)$. For the integrand dimensions we consider the cases $d=2,3,4$.

For generating the results, we made multiple identical integration runs and report the average uncertainty. It shall be noted that we observed a rare effect in the four-dimensional case in the low number of shots region, which we attribute to insufficient coverage of the Important Region and the imperfect training. In these rare runs, the variance would be getting considerably inflated. This was a consequence of not adequate sampling of one of the peaks in the extended boundary region. Thus, a large tile would be generated and evaluated with a single sample, which could fall to a badly covered part in proximity to the Important Region. To counteract this effect, we use a defensive IS mixture, to stabilize such a behavior, a strategy that is considered unbiased~\cite{hesterberg1995weighted}. The mixture we have used is to sample $90 \%$ of the points from the quantum-generated proposal PDF, and the remaining $10 \%$ from a uniform PDF that are distributed in the grid after the QAIS proposal has generated it.  

Figure~\ref{fig:4Dminimal} illustrates the results, where we compare QAIS with VEGAS for two-, three- and four-dimensional integrals in the form given by \Eq{eq:benchmark_int}, across the full sampling range usually examined in MC integration. The main observation is that every QAIS curve stays well below its VEGAS counterpart over the entire range of sample budgets considered, demonstrating a robust accuracy gain that persists regardless of the dimensionality or the sample budget.

Another observation is that QAIS exhibits an increase on the uncertainty at roughly between $10^4$ to $10^5$ shots. The effect is clearest in the four and three-dimensional curves. This occurs because, in this budget range, the proposal PDF has still not covered and is still discovering thin boundary layers around the Important Region, in which the integrand is very small, but not totally exponentially vanishing. Thus, some regions are wrongly categorized in the non-Important Region and generate weights that do not reflect their actual contribution.

To understand this effect, it is crucial to note that if we omit the debiasing step and retain the underestimated value with its corresponding standard deviation, the resulting uncertainty curves would be as flat as those of VEGAS. Nonetheless, the modest rise in uncertainty is a small price for obtaining unbiased results. Especially while considering that the extra uncertainty becomes insignificant above $10^5$ shots.  If sub-percent precision were needed at smaller budgets, a refined Tiling scheme that allocates extra samples to the boundary zones could suppress this effect without compromising unbiasedness.

On the overall cross-dimensional performance, based on our observations from Fig.~\ref{fig:4Dminimal}, the decisive metric is the Kullback-Leibler (KL) divergence. The two and three-dimensional proposals have divergences that are very close and their uncertainty difference is negligible above $10^5$ shots. This behavior aligns with our expectations, according to the form of the IS variance in~\Eq{eq:variance_qais}, since the statistical error is governed by the proposal PDF's quality rather than by the dimension of the integral. In the four-dimensional case, the best proposal we obtained with COBYLA had a comparatively larger divergence ($D_{\rm KL}\approx 0.21$), shifting its uncertainty curve upwards. Nevertheless, it reaches the accuracy and surpasses the two-dimensional VEGAS case in the sub-percentage accuracy region, hinting that even a non-perfect KL divergence can generate precise results. Ultimately, in the asymptotic region it nearly matches the precision of the other two curves. 

To motivate the general scaling to higher dimensions, it is important to note that for a number of non-mutually covered peaks $p$, the number of phantom peaks is $p^d-p$. Then, only the fraction $p/p^d$ of all samples reaches the true peaks, the MC standard deviation grows as $\sigma \propto p^{\frac{d-1}{2}}/ \sqrt{N}$. Consequently, if a quantum-generated proposal PDF keeps the KL divergence low with a computationally competitive state preparation, even as the number of dimension grows, it will reduce the sample budget needed for a given precision in a scalable way and thus achieve significant improvements in accuracy.

Finally, we quantify the discretization resolution we expect in terms of qubits per dimension. For moderately sharp targets such as the 3-peak benchmark, a 5 qubit per dimension discretization is sufficient to capture the main structure and yields stable variance. Additionally, the pentagon integral provides an example of a substantially sharper HEP motivated target. In that case, discretizations with 8-9 qubits per dimension yield good performance, whereas lower resolutions lead to a larger variance. This discretization is necessary in only one of the 3 dimensions, in which the sharp structure lies, while the rest can be kept in coarser discretizations. Moreover, increasing the number of qubits increases the number of grid points exponentially, so the resolution can be refined systematically by adding only a small number of qubits per dimension, making the discretization a controllable resource.

For HEP predictions of decay rates and cross sections, each extra particle in the final state contributes with 3 integration variables due to the phase space. In the LTD framework, each loop contributes with another 3 integration variables corresponding to the loop tree-momentum. It should be noted that phase-space and loop integrations are considered together in LTD. A two-body decay rate at NLO requires two independent integration variables (See Ref.~\cite{deLejarza:2024scm}), five independent variables at NNLO and eight at N3LO. For a $2\to 2$ scattering, we depart from four independent integration variables at NLO, with two extra variables at a hadron collider due to integration over parton densities. At $e^+e^-$ colliders, seven variables are needed at NNLO for $2\to 2$ scattering and nine at a hadron collider. If we consider a $1\to 3$ decay or a $2\to 3$ scattering, we have to add three extra variables. Thus, state-of-the-art applications require to begin at the range of $7-12$ dimensional integrals that would fall in the range of $35-120$ qubits, with the lower end of these ranges being more realistic, since the upper end would be needed only for very sharp structures that require very fine discretization in every single direction.

\subsection*{Trainability for PQC State Preparation}\label{sec:4.3}

%%%%%%%%%%%%%%%%%%%%%%%%%%%%%%
\begin{figure*}[t!]
\begin{center} 
\captionsetup{justification=Justified, singlelinecheck=off} 
\includegraphics[width=0.95\textwidth]{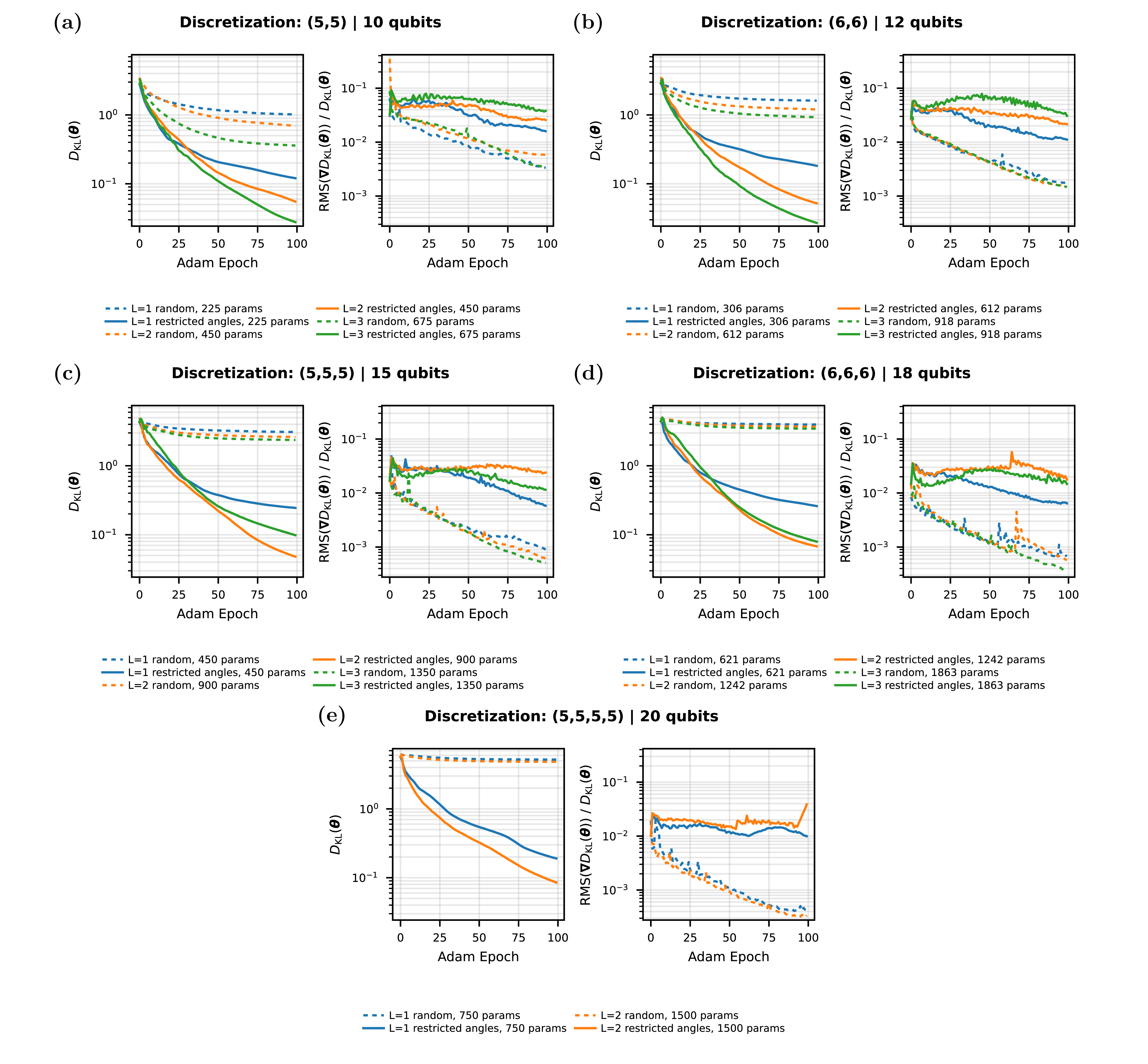} 
\caption{\textbf{Trainability across discretizations and integration dimensions.}  { \bf (a)-(e) } show trainability indicators for the Parameterized Quantum Circuit’s state preparation on the multi-peak benchmark described in the Results subsection \hyperref[sec:4.2.2]{Multi-peak Benchmark Integrals}, across discretizations from 10 to 20 qubits and 2-4 dimensional integrals. For each discretization, the Kullback-Leibler (KL) divergence evolution by the Adam epoch and the relative gradient scale $\mathrm{RMS}(\nabla D_{\mathrm{KL}})/D_{\mathrm{KL}}$ are shown. Solid lines correspond to restricted-angle initialization, while dashed lines correspond to random initialization. The label $L=k$ denotes $k$ layers of the PQC sequence $U^{(Z)} \to R_Z \to R_X \to R_Y \to U^{(Y)} \to R_Z \to R_X \to R_Y \to U^{(X)} \to R_Z \to R_X \to R_Y$. The corresponding numbers of trainable parameters are also shown.
}
\label{fig:grads}
\end{center}
\end{figure*}
%%%%%%%%%%%%%%%%%%%%%%%%%%%%%

Trainability is a central challenge in scaling variational quantum algorithms to larger systems. In highly expressive PQCs, gradient-based optimization can become ineffective due to vanishing gradients, a phenomenon known as barren plateaus~\cite{McClean_2018, Larocca_2025,Holmes_2022}. Additionally, the optimization landscape can be strongly non-convex, with multiple local minima~\cite{you2021exponentiallylocalminimaquantum,Anschuetz_2022}. These effects are particularly relevant for global objective functions, where the typical gradient components, that have a zero expectation value by construction, also get an exponentially small variance in the number of qubits, making meaningful parameter updates increasingly unlikely as system size grows.

Motivated by these observations, we include explicit optimization diagnostics for the PQC state-preparation stage of QAIS. The goal is to probe how the optimization signal remains under the chosen architecture and optimizer, and to identify early signs of stagnation as the qubit count of the discretization, and circuit size increase.

To mitigate trainability issues, we use a structured initialization rather than fully random angles, which are known to exacerbate barren plateau behavior in expressive PQCs. The details of the implementation are provided in the Methods subsection \hyperref[sec:trainability_methods]{Trainability Diagnostics}.

Figure~\ref{fig:grads} reports representative runs for the multi-peaked benchmarks discussed in the Results subsection \hyperref[sec:4.2.2]{Multi-peak Benchmark Integrals} across multiple integration dimensions and discretizations (10 to 20 qubits). For each setting, we present the loss $D_{\mathrm{KL}}$ and the relative gradient scale $\mathrm{RMS} \left(\nabla D_{\mathrm{KL}}(\boldsymbol{\theta})\right)/D_{\mathrm{KL}}(\boldsymbol{\theta})$ as a function of Adam iterations, under identical architectures and optimizer settings, comparing the random initialization to the restricted-angle initialization.

Across the discretizations shown, random initialization exhibits loss stagnation in $D_{\mathrm{KL}}$ and systematic deterioration of the optimization signal as the training proceeds and as size of the problem increases. In contrast, the targeted initialization around the maximally superposed state is able to train toward values of the KL divergence that we consider successfully trained, and it does so with a steady optimization signal over the reported iterations. We further observe that, in several settings, increasing the PQC's expressivity can lead to lower achieved minima in $D_{\mathrm{KL}}$ under restricted-angle initialization. This behavior is consistent with a regime in which additional expressivity can be exploited by the optimizer when gradients remains non-negligible.

While these diagnostics do not constitute a full scaling analysis, they provide evidence that the state preparation through a QCBM exhibits a certain stability in the proof-of-principle regimes studied here when combined with a targeted initialization, motivating further investigation of structured initialization and training strategies for larger size problems. 

Moreover, we emphasize that QCBMs constitute a convenient, but neither unique nor necessary, approach to do state preparation and subsequently apply the QAIS integral estimation framework. One of the main strengths of QAIS is its independence from the method employed to prepare the initial state, allowing alternative strategies that are less susceptible to trainability issues to be equally viable. In particular, approaches based on tensor-networks \cite{ Melnikov:2023duq,ballarin2025efficientquantumstatepreparation}, as well as other state preparation alternatives which do not necessarily rely on variational training for loading discretized representations of continuous functions \cite{akhalwaya2023modularenginequantummonte,O_apos_Brien_2025}, represent a promising direction that we leave for future work.

%%%%%%%%%%%%%%%%%%%%%%%%%%%%%%%%%%%%%%%%%%%%%%%%%%%%%%%%%%%%%%%%%%%%%%%%%%%%%%
\section*{Discussion}

We have presented a hybrid (quantum-classical) Monte Carlo integration method that performs Adaptive Importance Sampling~(AIS) using a Parametrized Quantum Circuit~(PQC). Inspired by the success of classical AIS integrators, such as VEGAS, we aim to address their limitations and identified the two main directions in which quantum computing can contribute.

Firstly, the standard approach to numerical integration using AIS involves discretizing the integration domain into a multidimensional grid. The grid's size grows exponentially with the number of dimensions, and manipulating a Probability Density Function~(PDF) in the integration domain through such a grid soon becomes computationally prohibitive. To overcome these drawbacks, we encode the grid directly into an $n$-qubit Hilbert space and manipulate the PDF via a strongly entangled PQC. We employ generative modeling to train and obtain an optimal PQC whose PDF accurately reflects the underlying structure of the target integrand.

Second, although Importance Sampling is an exceptionally effective variance reduction framework, its performance depends critically on the proposal PDF. By shaping this probability distribution with a highly expressive entangled PQC, we efficiently sample from complex non-separable PDFs that capture directly the correlations in the multidimensional space. This strategy enhances the performance of Importance Sampling.

For the demonstration of the performance of the method, we applied QAIS to integrate two challenging types of functions and compared the results with VEGAS, which serves as our reference. The comparison focuses on the accuracy that each method's optimal proposal PDF achieves. The functions considered are a sharply-peaked single-mode one-loop Feynman integrand, and a set of multi-modal benchmark integrands from two to four dimensions. For the first case, the two methods performed comparably overall. For the multi-modal functions, VEGAS allocated a substantial amount of samples in non-Important Regions, especially as the dimension increased, due to its separable proposal PDF, whereas QAIS consistently maintained impressive accuracy.

As a future work, we intend to make our integration workflow more competitive by refining the current optimization procedure. While QAIS already represents richer PDFs, a reliable optimization routine, especially as robust as the one of VEGAS, remains challenging, as in many current Quantum Machine Learning approaches. We therefore plan to formalize the optimization procedure in detail with more specific PQC architectures, combining them with initialization of the optimization process with easily loadable approximations of the target functions and other established heuristics. We also note that a comparison with classical ML-based integrators is within the scope of future studies. In addition, a shot-noise aware analysis is needed to assess the impact of finite shot sampling on the state preparation, which we view as the current priority. The impact of hardware noise on the state-preparation stage and on the final estimator variance is also left for future work, together with the associated trade-offs in PQC size, shot budget, and mitigation overhead in a NISQ oriented or early-fault tolerant setting.

\section*{Methods}

\subsection*{Monte Carlo integration, Importance Sampling and VEGAS}
\label{sec:sec2}
%%%%%%%%%%%%%%%%%%%%%%%%%%%%%%%%%%%%%%%%%%%%%%%%%%%%%%
%%%%%%%%%%%%%%%%%%%%%%%%%%%%%%%%%%%%%%%%%%%%%%%%%%%%%%

Monte Carlo (MC) integration has long been the most widely accepted method of estimating multidimensional integrals. In its most basic formulation, the MC estimation of an integral is obtained by randomly sampling points uniformly across the integration domain and evaluating the integrand at these points. In particular, consider the integral 
\begin{equation}
    I = \int_{\Omega} f({\bf x}) d{\bf x}~,
    \label{eq:target}
\end{equation}
where {\bf x} is a vector in $\mathbb{R}^d$,  $ \Omega \subset \mathbb{R}^d $ corresponds to the integration domain, and the function $f : \Omega \to \mathbb{R} $ is the integrand. By drawing a set of $N$ independent and identically distributed random samples ${\{\bf x}_1, \ldots, {\bf x}_N\}$, one obtains the MC estimator 
\begin{equation}
    \hat{I}_N^{\rm (MC)} = \frac{|\Omega|}{N}\sum_{i=1}^N f({\bf x}_i)= |\Omega| \langle f \rangle~,
\end{equation}
with the corresponding variance of:
\begin{equation}
\left(\hat{\sigma}_N^{(\rm MC)}\right)^2= \frac{|\Omega|^2}{N - 1}\left(\frac{1}{N}\sum_{i=1}^{N}f(\mathbf{x}_i)^2 - \langle f\rangle^2\right)~.
\end{equation}

In general, the power of this method lies in the fact that it imposes minimal demands on the integrand such as the function does not need to be smooth or analytic, demands that in many practical applications, such as in HEP, are often violated. 

Importance Sampling (IS) is a statistical technique used to estimate the value of integrals, especially for complex and high-dimensional functions. The objective of IS is to significantly reduce the variance of the estimate by sampling from a carefully chosen PDF. In IS, the integral in \Eq{eq:target} is reformulated as:
\begin{equation}
    I=\int_\Omega \frac{f({\bf x})}{q(\bf{x})} q({\bf x}) d{\bf x}~, 
\end{equation}
where $q({\bf x})$ is the proposal PDF that is both computationally efficient to sample from and closely resembles $f({\bf x})$. In this setup, and given a set of samples $\{{\bf x}_1,\dots, {\bf x}_N \}$ drawn from $q({\bf x})$, the MC estimator becomes:
\begin{equation}\label{eq:IS_estimator}
        \hat{I}_N^{\rm (IS)} = \frac{1}{N} \sum_{i=1}^N \frac{f({\bf x}_i)}{q({\bf x}_i)}~.
\end{equation}

The variance of the IS estimator is written as:
\begin{equation}\label{eq:variance_importance_sampling}
\left(\hat{\sigma}_N^{(\mathrm{IS})}\right)^2 = \frac{1}{N - 1}\left(\frac{1}{N}\sum_{i=1}^{N}\frac{f(\mathbf{x}_i)^2}{q(\mathbf{x}_i)^2}- \bigl(\hat{I}_N^{(\mathrm{IS})}\bigr)^2\right)~.
\end{equation}
The precision of the estimator is highly dependent on the choice of $q({\bf x})$. If there is a large discrepancy between $f({\bf x})$ and $q({\bf x})$, such that $q({\bf x})$ does not adequately capture the behavior of $f({\bf x})$, then this discrepancy will propagate to the overall variance, practically increasing it.

%%%%%%%%%%%%%%%%%%%%%%%%%%%%%%%%%%%%%%%%%%%%%%%%%%%%%%%%%%%%%%%%%%%%%%%%%%%%%%%%%%%%%%%%%%%%%%%%%%%%
%\subsection{VEGAS algorithm in HEP}

The classical  version of VEGAS~\cite{Lepage:1977sw,Lepage:123074} implements an Adaptive Importance Sampling~(AIS) strategy to estimate multidimensional integrals. VEGAS reformulates the target integral in \Eq{eq:target} as an integral over the unit hypercube,
\begin{equation}
  I=\int_{[0,1]^d} J(\mathbf{y})f\bigl(\mathbf{x}(\mathbf{y})\bigr) d\mathbf{y},
\end{equation}
by introducing a change of variables $\mathbf{x}=\mathbf{x}(\mathbf{y})$ with Jacobian $J(\mathbf{y})=\bigl|\partial\mathbf{x}/\partial\mathbf{y}\bigr|$, chosen to flatten the peaks of $f({ x})$ and thus minimize the MC variance. The points $\{\mathbf{y}_i\}$ are sample points drawn from a $d$-dimensional uniform PDF in the $\mathbf{y}$-space, i.e. $[0,1]^d$, and then are mapped to the original $\mathbf{x}$-space.

On each dimension, each axis is divided into $N_g$ intervals of varying widths $\{\Delta { x}_i\}_{i=0}^{N_g-1}$,
\begin{equation}
  { x}_0={ a},\quad { x}_i = { x}_{i-1} + \Delta { x}_{i-1},\quad { x}_{N_g}={ b},
\end{equation}
and a point from the ${ y}$-space is mapped to the ${ x}$-space by
% \begin{equation}
%   { x}({ y})= { x}_i + \Delta { x}_{i} \bigl({ y} N_g - i \bigr), 
%   \quad i =\lfloor y N_g\rfloor
%   \qquad (0\le y\le1) . 
% \end{equation}
\begin{equation}
\begin{split}
x(y) &= x_i + \Delta x_i \bigl(y N_g - i \bigr), \\
i &= \lfloor y N_g \rfloor, \qquad (0 \le y \le 1)
\end{split}
\end{equation}
with $i$ being the index of the grid cell, in which the point belongs to. The Jacobian is the step function
\begin{equation}
  J({ y})=N_g \Delta { x}_{i} , 
\end{equation}
so that with this transformation, a uniform draw in $y$-space concentrates samples in regions where the $\Delta {x}_i$ are the smallest. Thus, the objective becomes to use finer widths, where $|f({x})|$ is largest.

In $d$ dimensions this mapping is applied independently to every coordinate, giving $d$ one-dimensional grids, each with $N_g$ cells, instead of the $N_g^{d}$ cells of a full discretization. Because in this setting, the Jacobian factorizes as 
\begin{equation}\label{eq:projections}
J(\mathbf y) = \prod_{i=1}^d J_i(y_i),
\end{equation}
the induced proposal PDF is separable. This product form of the proposal PDF eliminates the exponential computational cost of sample allocation but also cannot capture correlation between different dimensions, thus in certain cases leads to misallocating samples, creating artificial structures (e.g. phantom peaks) or undersampling high-impact regions. This problem becomes amplified as the integrand dimension increases. 

With $N$ samples, the VEGAS estimator is
\begin{equation}
  \hat I_{N}^{\rm (VEGAS)} = \frac{1}{N}\sum_{i=1}^{N} J(\mathbf{y}_i) f\left(\mathbf{x}(\mathbf{y}_i)\right),
\end{equation}
whose variance is:
\begin{equation}\label{eq:variance_vegas}
\begin{split}
\left(\hat{\sigma}_N^{\mathrm{(VEGAS)}}\right)^2 &= \frac{1}{N-1} \Biggl[ \frac{1}{N}\sum_{k=1}^{N} J^{2}(\mathbf{y}_k)\, f^{2}\bigl(\mathbf{x}(\mathbf{y}_k)\bigr)\\
&\hspace{7.0em} - \left(\hat{I}_N^{\mathrm{(VEGAS)}}\right)^2 \Biggr].
\end{split}
\end{equation}
To balance this variance VEGAS adapts each one-dimensional grid after every iteration. Let $n_i$ be the number of samples that fall in cell $i$. The algorithm defines the variable, 
\begin{equation}
D_i = \frac{1}{n_i}\sum_{x(y)\in\Delta x_i} J^{2}(y)f^{2}\bigl(x(y)\bigr).
\end{equation}

Next, it smooths and compresses the $D_i$, to a nonlinear, tunable parameter controlled, smoothed and compressed value derived from its current value, its two neighbours, and the global sum. With the $D_i$ being smoothed and compressed, the algorithm chooses the new cell's boundaries $\{ \Delta x'_i \}$, so that each updated bin contains exactly $\bar{ D } = \frac{1}  {N_g}\sum_{j=0}^{N_g-1} D_j $ thereby forcing each new bin to contribute equally to the total variance. Practically, one starts at the left edge of an axis, accumulates the $D_i$ until the running sum reaches the target $\bar{D}$ , then inserts a new boundary at that point. After setting the new boundary, any excess $D_i$ transfers to the next bin, and continues from left to right until the whole axis is re-discretized. This is done separately, dimension by dimension. In each iteration, new samples are generated, and thus new $D_i$ are used on the updated grid.

Since VEGAS is adaptive, each pass produces an estimate $I_j$ with variance $\sigma_j^2$, and after~$j$ adaptations the combined result is

\begin{equation}\label{eq:finalvegasestimateformulas}
\begin{split}
{\bar I}_N^{\rm (VEGAS)}&=\frac{\sum_{j} \hat I_{N,j}^{\rm (VEGAS)}/\left(\hat \sigma_{N,j}^{\rm (VEGAS)}\right)^{2}}{\sum_{j}1/\left(\hat \sigma_{N,j}^{\rm (VEGAS)}\right)^2}, \\
{\bar \sigma }_N^{\rm (VEGAS)}&=\left(\sum_{j}1/\left(\hat \sigma_{N,j}^{\rm (VEGAS)}\right)^2\right)^{-1/2},
\end{split}
\end{equation}
which carries a smaller uncertainty than any single iteration. Because the weights are inverse variances, the best adaptations dominate the average. In this work, we compare with the original importance sampling VEGAS as implemented in the standard FORTRAN code~\cite{Lepage:123074}. Finally, stratified sampling extensions~\cite{Lepage:2020tgj} can further lower errors by exploiting more elaborate statistical techniques but still the general scaling to multidimensional integrals suffers from the curse of dimensionality.

\subsection{Constructing the Grid with a Parametrized Quantum Circuit}\label{sec3.1}

Our objective is to perform IS for numerical integration using an optimal PDF, which is strategically adapted by a quantum protocol on a grid. The first step in this direction is to define the grid through a quantum state. For an $n$-qubit system, the underlying Hilbert space, is given by:
\begin{equation} 
\mathcal{H} =\underbrace{\mathbb{C}^2 \otimes \mathbb{C}^2 \otimes \cdots \otimes \mathbb{C}^2}_{n \otimes }=\left(\mathbb{C}^2\right)^{\otimes n}~,
\end{equation}
where each qubit is associated with a two-dimensional complex vector space $\mathbb{C}^2$. The dimension of the space is $2^n$. For a $d$-dimensional integral, each dimension is encoded using $q_i$ qubits. Hence, the full size of the system is $n=\sum_{i=1}^d q_i$, and the reformulation of the Hilbert space is:
\begin{equation}
\mathcal{H}=\left(\mathbb{C}^2\right)^{\otimes n}=\bigotimes_{i=1}^d \left(\mathbb{C}^2\right)^{\otimes q_i}~.
\end{equation}

The grid's resolution of each dimension is determined by the number of qubits used to represent this dimension. Using more qubits per dimension will reveal finer details of the integrand structure. The grid coordinates are generated through the Big-Endian Encoding (BEE), a most-significant-bit-first convention to map bitstrings to integer grid indices, which is  consistent with the computer-science Big-Endian convention~\cite{1667115}. This approach is demonstrated explicitly in a 1D quantum encoding example in Ref.~\cite{Iaconis_2024}. For a $d$-dimensional domain with integration bounds $\mathbf{a}=(a_1,\dots,a_d)$ and $\mathbf{b}=(b_1,\dots,b_d)$, each dimension $i$ is represented by $q_i$ qubits, corresponding to $2^{q_i}$ intervals of length
\begin{equation}\label{eq:interv_size}
  \Omega_i = \frac{b_i-a_i}{2^{q_i}}~.
\end{equation}

A computational basis state $\ket{j_1}\otimes\cdots\otimes\ket{j_d}$ with $j_i \in \{0,\dots,2^{q_i}-1\}$ corresponds to and labels the grid cell, with boundaries:
\begin{equation}
  \prod_{i=1}^{d}\bigl[a_i + j_i\Omega_i, a_i + (j_i+1)\Omega_i\bigr]~.
\end{equation}
For a quantum system, in a general state  
\begin{equation}
  \ket{\psi} = \sum_{j_1=0}^{2^{q_1}-1} \cdots \sum_{j_d=0}^{2^{q_d}-1} c_{(j_1,\dots,j_d)} \ket{j_1}\otimes\cdots\otimes\ket{j_d}~,
\end{equation}
a $Z$-basis measurement returns the cell $(j_1,\dots,j_d)$ with probability $|c_{(j_1,\dots,j_d)}|^{2}$. By interpreting these probabilities as bin heights we construct the piecewise PDF in the continuous space as:
\begin{equation}
\begin{split}
q(\mathbf{x})&=
\sum_{j_1,\dots,j_d} |c_{(j_1,\dots,j_d)}|^{2}
\prod_{i=1}^{d} \frac{1}{\Omega_i}
\\
&\quad\quad \times \Bigl[ \Theta \bigl(x_i-(a_i+j_i\Omega_i)\bigr)
\\
&\qquad \qquad \qquad -\Theta \bigl(x_i-(a_i+(j_i+1)\Omega_i)\bigr) \Bigr]~.
\end{split}
\label{eq:pdf-big-endian}
\end{equation}
Equation~\eqref{eq:pdf-big-endian} defines a piecewise‐uniform PDF in the integration domain $\Omega$. Because a $Z$‐basis measurement samples exactly one of the $2^{n}$ basis states, the shot record $\{k_{(j_1,\dots,j_d)}\} $, follows a multinomial distribution with probabilities $p_{(j_1,\dots,j_d)}=|c_{(j_1,\dots,j_d)}|^{2}$.  For a fixed PQC at $N$ measurements, the maximum‐likelihood estimator is simply the empirical frequency, $ {\hat p_{(j_1,\dots,j_d)} } = k_{(j_1,\dots,j_d)}/N$. Thus, the BEE construction provides a direct way to construct the full $2^{n}$-cell grid, while the resolution along any dimension can be refined simply by adding qubits to that dimension.

The core idea of QAIS is the implementation of AIS directly with a PQC. This allows us to operate in a computationally efficient way across the entire Hilbert space $\bigl(\mathbb{C}^2\bigr)^{\otimes n}$. Through entanglement, quantum gates affect qubits that belong to cross-dimensional parts of the Hilbert space, enabling correlations that go beyond the axis-only separability assumption, thus letting samples to flow from any part of the integration domain to any other. With this approach, we avoid the approximation typically required by the exponential growth of the integration domain's grid size. In particular, we overcome the separability assumption of VEGAS, which intuitively can be thought as restricting to a separable quantum state $\ket{\psi}=\ket{\psi_1} \otimes \ket{\psi_2} \otimes \cdots \otimes \ket{\psi_d}$. 

Another difference at the encoding level is that QAIS allocates samples by varying the bin heights of the different states through the square amplitudes $|c_{(j_1,\dots,j_d)}|^2$ and increases the resolution by increasing the number of qubits. VEGAS concentrates samples in regions of importance by deforming the grid, while keeping the bin height equal on all of the grid cells.

%%%%%%%%%%%%%%%%%%%%%%%%%%%%%%%%%%%%%%%%%%%%%%%%%%%%%%%%%%%%%%%%%%%%%%%%%%%%%%%%%%%%%%
\subsection{Quantum Importance Sampling Statistical Framework}\label{subsec:3.2}

Having defined the PDF on the integration domain through BEE, we are now in a position to construct the statistical framework for performing quantum IS. For this purpose, we clarify how the quantum basis states samples relate to the underlying integration domain. Let $\Omega$ be our full discretized integration domain, corresponding to $2^n$ distinct grid cells. We label these cells as $\{ \Omega^{(i)} \}_{i=1}^{2^n} $, where each corresponds to a particular basis-state $\ket{i}$ in the computational $Z$-basis. Each state $\ket{i}$ is mapped onto a specific, $d$-dimensional grid interval, according to the notation introduced in the Methods subsection \hyperref[{sec3.1}]{Constructing the Grid with a Parametrized Quantum Circuit}. Nevertheless, in a linearized setting, the full integration domain is expressed as
\begin{equation}
\label{eq:domain-partition}
    \Omega = \bigcup_{i=1}^{2^n} \Omega^{(i)}, 
    \quad \Omega^{(i)} \cap \Omega^{(j)} = \varnothing \quad (\forall ~ i \neq j)~.
\end{equation}

Each grid cell $\Omega^{(i)}$ forms a hyper-rectangle, defined explicitly by intervals determined through the discretization recipe described by~\Eq{eq:interv_size}. Specifically, the grid cell volumes are given by
\begin{equation}
|\Omega^{(i)}|=\prod_{k=1}^d \Omega_k=\prod_{k=1}^d \frac{b_k-a_k}{2^{q_k}}~.
\end{equation}
These volumes are constant for all of the $2^n$ grid cells, of the integration space, given the equidistant discretization per dimension. 

To proceed, we assume state preparation as granted, thus we have a PQC trained to produce a quantum state whose measured PDF approximates the target function of interest in the grid. After sampling from the PQC, we denote with $N_i$ the number of occurrences of the state $\ket{i}$. From a total of $N$ measurements, drawn from the quantum proposal PDF, we obtain $N_i$ samples within each cell $\Omega^{(i)}$. Within each cell, these $N_i$ samples are further distributed to the continuous intervals by generating $N_i$ uniform random points. In our implementation, we draw each cell’s samples using quasi-random points, specifically, from a Sobol sequence~\cite{SOBOL196786}. This is fully compatible with the discretization and sample allocation by a PQC while it reduces the error further. However, for comparability, we apply and report the standard variance formula. The local integral of the target function $f({\bf x})$ in the cell $\ket{i}$ is 
\begin{equation}
\label{eq:subdomain-integral}
    I_i=\int_{\Omega^{(i)}} f({\bf x}) d{\bf x}~.
\end{equation}
Hence, the local MC estimator in $\Omega^{(i)}$ becomes
\begin{equation}
\label{eq:local-mc-estimator}
    \hat{I}_i =\frac{|\Omega^{(i)}| }{N_i}\sum_{j=1}^{N_i} f\left({\bf x}_j^{(i)}\right)~,
\end{equation}
where the index $i$ corresponds to the $i$-th grid cell, and the index $j$ to the $j$-th random point. Since $N_i$ is the number of times the PQC output fell into cell $i$, we define the weight of this grid cell as:
\begin{equation}
\label{eq:cell-weight}
w^{(i)}=\frac{|\Omega^{(i)}| N}{  N_i}~.
\end{equation}
Based on these definitions and by considering the form of the IS estimator of \Eq{eq:IS_estimator} and the form of the proposal PDF of \Eq{eq:pdf-big-endian}, the total estimator for the integral over $\Omega$ is:
\begin{equation}
\label{eq:full-estimator}
    \hat{I}_N^{\rm (QAIS)}= \frac{1}{N} \sum_{j=1}^N w^{(i)}f\left({\bf x}_j^{(i)}\right).
\end{equation}
The variance of the integral estimate is:
% \begin{equation}\label{eq:variance_qais}
% \left(\hat \sigma^{\rm (QAIS)}_N \right)^2=\frac{1}{N-1} \left(\frac{1}{N} \sum_{j=1}^N \Bigl[w^{(i)} f({\bf x}_j^{(i)})\Bigr]^2- \left(\hat{I}_N^{\rm (QAIS)}\right)^2 \right)~.
% \end{equation}
\begin{equation}\label{eq:variance_qais}
\begin{split}
\left(\hat \sigma^{\rm (QAIS)}_N \right)^2&=\frac{1}{N-1} \Biggl(\frac{1}{N} \sum_{j=1}^N \Bigl[w^{(i)} f({\bf x}_j^{(i)})\Bigr]^2\\
&\hspace{7.0em}-\left(\hat{I}_N^{\rm (QAIS)}\right)^2 \Biggr)~.
\end{split}
\end{equation}
While an IS estimator is, in principle, unbiased, there is a subtle but crucial consideration in practice that becomes highly relevant within a quantum computational setting. For quantum computing to offer a meaningful advantage, it is essential that the PQC focuses sampling on a small subset of the entire $2^n$-dimensional Hilbert space specifically those subdomains that contribute significantly to the integral. Typically, integrals over high-dimensional spaces are dominated by a relatively limited region of fluctuating behavior or sharp peaks, while most of the integration domain contributes smoothly or negligibly. This behavior is precisely what motivates the use of IS in a quantum computational framework, since it exactly aligns with the constraints and objectives of both the statistical and the quantum computational setting. However, this advantage also implies that a portion of the integration domain associated to rarely or never observed computational basis states will be overlooked. These regions remain unobserved not because their contribution is strictly zero, but rather because their probability, as defined through the PQC, is too small to produce occurrences in the finite number of quantum measurements performed.  

After performing $N$ measurements, we observe a subset of the Hilbert space elements or grid cells, $\Omega^- \subset \Omega$. The remaining states (or grid cells), $\Omega \setminus \Omega^-$, are effectively unseen. In this setup, the estimator of \Eq{eq:full-estimator} can be re-expressed as:
\begin{equation}
\hat{I}^{\rm (QAIS,obs)}_N = \frac{1}{N}\sum_{i\in \Omega^-} w^{(i)} \sum_{j=1}^{N_i} f({\bf x}^{(i)}_{j}).
\end{equation}
The missing contribution from $\Omega \setminus \Omega^-$ introduces a systematic bias to the integral estimation, leading to an underestimation in its absolute value. This bias can be explicitly evaluated as 
\begin{equation}
\label{equation:bias}
\text{Bias}_{\Omega \setminus \Omega^- }=\mathbb{E}\left[\hat{I}^{\rm (QAIS,obs)}_N\right]-I=-\sum_{i \in \Omega \setminus \Omega^-} I_i ~,
\end{equation}
where $I$ is the true integral over the complete domain $\Omega$ and $\hat{I}_{\text{obs}}$ is the estimator of the integral, based solely on the observed outcomes from performing $N$ measurements on the PQC.

%%%%%%%%%%%%%%%%%%%%%%%%%%%%%%%%%%%%%%%%%%%%%%%%%%%%%%%%%%%%%%%%%%%%%%%%%%%%%%%%
\subsection{Debiasing Strategy using a Tiling Algorithm}\label{subsec:3.3}

To quantify and correct this bias, we conceptually partition the integration domain into three regions:
\begin{itemize}
\item The \textbf{Important Region} $\Omega_{I}\subset\Omega^-$, containing basis states with significant integrand values and high sampling probability.
\item The \textbf{Boundary Region} $\Omega_{B}\subset\Omega^-$, containing adjacent states that surround the Important Region.
\item The \textbf{Noise Region} $\Omega_{N}\subset\Omega^-$, consisting of states that are sporadically observed and vary, as a set, across different measurement runs due to shot noise. These states contain low but non-zero probability and appear in isolation, without adjacent measured states.
\end{itemize}
The union of these regions is $\Omega^-= \Omega_{I} \cup \Omega_{B} \cup \Omega_{N}$. For $M$ measured states after $N$ quantum measurements, the separation defined above identifies the $N_I$ states belonging to the core of the Important Region and the $N_{B\cup N} = N_B + N_N$ states in the Boundary and Noise Regions, respectively, such that $M = N_I + N_{B\cup N}$.

Our strategy to correct the bias introduced by the unobserved states is to assume that the observed Boundary and Noise Regions are representative samples from the otherwise unobserved region $\Omega\setminus\Omega^-$. This is a heuristic assumption introduced to ensure that every otherwise empty tile is populated without requiring extra sampling. Essentially, we build a representative approximation for what we call the non-Important Region $\Omega_{N-I}= ( \Omega \setminus \Omega^- ) \cup \Omega_{B} \cup \Omega_{N} $ and thus reconstruct the PDF in the whole integration domain $\Omega$. Due to our pre-training assumptions, we expect that a very precise sampling of the non-Important Region is unnecessary, as their overall contribution to the integral and consequently the corresponding uncertainty are minor. Additionally, our main focus is to construct the non-Important Regions computationally efficiently, rather than generating a very detailed representation of the substructures that exist within them.

In practice, for each of the $N_{B\cup N}$ cells in the Boundary and Noise Regions, we construct an enlarged hyper-rectangular domain $\{ \tilde{\Omega}^{(k)} \}_{k=1}^{N_{B\cup N}}$. Each enlarged domain $\tilde{\Omega}^{(k)} $ contains exactly one measured boundary or noise cell along with its adjacent unobserved states, ensuring a one-to-one correspondence with $\Omega_B\cup\Omega_N$, and a number of hyper-rectangles bounded from above by $M$. The previously missing integral contribution associated with the unobserved cells, is explicitly taken into account by the redistribution of samples from $\Omega_B \cup \Omega_N$ to these enlarged sub-domains, while ensuring no extra sampling is required. The integral estimator, in this corrected setup is:
\begin{equation}\label{eq:stratified_estimator}
\begin{split}
\hat{I}^{\rm (QAIS)}_N &= \frac{1}{N} \sum_{i \in \Omega_{I}} w^{(i)} \sum_{j=1}^{N_i} f({\bf x}^{(i)}_{j})\\
&\hspace{4.0em}+\frac{1}{N}  \sum_{k\in \Omega_{N-I}} \tilde{w}^{(k)}  \sum_{j=1}^{\tilde{N}_k} f({\bf x}^{(k)}_{j})~,
\end{split}
\end{equation}
where the adjusted weights $\tilde{w}^{(k)}$ take into account the shapes of the new enlarged hyper-rectangular domains and $\tilde{N}_k$ refers to the number of random samples drawn within each enlarged domain $\tilde{\Omega}_k$. Because the debiasing step essentially groups the cells, re-labels them and repositions some of the existing $N$ samples, the estimator's formula of \Eq{eq:stratified_estimator} is equivalent to \Eq{eq:full-estimator}, with the variance given by \Eq{eq:variance_qais}, provided that the adjusted weights are used where it is appropriate. Through this construction, we explicitly include all previously neglected states, eliminating missing regions and ensuring that the estimator is unbiased.

In practice, for constructing the non-Important Region, we employ a Tiling algorithm, which is described in detail in the Supplementary Methods. This algorithm generates a small number of contiguous, non-overlapping hyper-rectangles that collectively cover every previously unobserved grid cell, a process that corresponds to a full tiling of the $d$-dimensional integration domain. For the construction of such an algorithm, it is important to have a computational cost that scales polynomially with the number of observed states $M$, the dimension of the integration space $d$, and  the number of qubits used for discretization $n$. It is an explicit and crucial constraint for the Tiling algorithm not to scale with the number of unobserved grid cells, since, in this framework, this is an exponential sized quantity.

The Tiling algorithm we compose is based on the delta encoding~\cite{tridgell1996rsync,10.1145/263105.263162}. We compress each contiguous block of unmeasured outcomes between two observed states into a single gap, storing it by its length and its lower binary boundary so that we can later sample uniformly within it. To cover the full integration space, these gap records are converted into hyper-rectangle intervals by a routine similar to the greedy-meshing heuristic well known in computer graphics~\cite{sykes2012meshing} and generalized here to arbitrary $d$-dimensional grids.

The procedure works by generating greedy expansions between blocks of different dimensions. By leveraging this approach, the algorithm skips large blocks of lower-ordered dimensions, when making expansions in a higher ordered dimension. In this way, it efficiently tiles arbitrarily sized gaps in the $d$-dimensional grid with hyper-rectangles, using at most $2(d-1)+1$ intervals per gap. This bound on the number of intervals is the key factor behind the suitable scaling of the Tiling algorithm. After accounting for all operations, even the arithmetic operations, the tile generation executes in time $\mathcal{O}(M  \, d^3 \, n^2)$. This, compared to the state-sorting cost $\mathcal{O}(n \, M \log M)$, is typically negligible. Therefore, the overhead coming from the Tiling procedure is essentially $\mathcal{O}(M \log M)$.

%%%%%%%%%%%%%%%%%%%%%%%%%%%%%%%%%%%%%%%%%%%%%%%%%%%%%%%%%%%%%%%%%%%%%%%%%%%%%%%%%%%%% 
\subsection{Structure and Optimization of the Parametrized Quantum Circuit}\label{sec:sec3.4}

In order to perform state preparation, we employ a QCBM to train the PQC of the proposal PDF. The target function $f(\mathbf{x})$ is generally defined over a continuous space and is discretized along with the integration domain into a grid, as discussed in the Methods. For training, we use the KL divergence \cite{Kullback:1951zyt} as a cost function to measure and minimize the distance between the target distribution and the proposal PDF. 

We use a discretized version of the KL divergence, which is defined as:
\begin{equation}
    D_{\text{KL}}(P \| Q) = \sum_{i=1}^{M} P(\Omega^{(i)}) \log \left( \frac{P(\Omega^{(i)})}{Q(\Omega^{(i)})} \right)~,
\end{equation}
where $P(\Omega^{(i)})$ denotes the probability corresponding to the target distribution at a single grid cell $\Omega^{(i)}$ ( or state $\ket{i}$ ) and $Q(\Omega^{(i)})$ denotes the probability of the proposal distribution at the grid cell to which $\Omega^{(i)}$ belongs.

To convert the continuous function $f(\mathbf{x})$, into a discrete PDF or a Probability Mass Function (PMF) suitable for computing the discretized KL-Divergence, we proceed as follows. We first compute the function at $N_i$ representative points \{$\mathbf{x}_i$\} within each observed grid cell $\Omega^{(i)}$, accounting in total to $N=\sum_{i=1}^M N_i$ function evaluations. Because the grid is uniform, every cell has the same volume, so common factors cancel in the normalization. Therefore, we construct the PMF by taking the simplified expressions, using $P(\Omega^{(i)})= f(\Omega^{(i)})/Z$ which is the probability of each grid cell, where $f(\Omega^{(i)})= 1/N_i \sum_{i=1}^{N_i} f(\mathbf{x}_i) $ is the within-cell function's average and $Z=\sum_{i=1}^{M} f( \Omega^{(i)} )$ the normalization constant. For the non-uniform grid case, the cell's estimate is computed as $f(\Omega^{(i)})= | \Omega^{(i)}| /N_i \sum_{i=1}^{N_i} f(\mathbf{x}_i) $, with the resulting change in $Z$. Ideally, for a better precision in the target PMF, the cell-wise sample count, $N_i$ would be updated after each run, by assigning more sample to high probability cells. This, combined with the monitoring of the within-cell variance in high impact region, provides a criterion that indicates whether the grid resolution must be increased by adding more qubits. However, in this work, for robustness and simplicity, we use a fixed $N_i$ per cell and cache all evaluations, while pre-defining the grid's size.

In a shot-based optimization, a proven strategy~\cite{rudolph2023trainability} is to adopt an implicit cost function such as the Maximum Mean Discrepancy (MMD) with a carefully chosen, qubit‐scaling Gaussian kernel, to avoid shot‐induced barren plateaus. However, in this work, rather than focusing on the intricacies of the training or optimization procedures, which are themselves very active areas of research in quantum machine learning~(QML) and quantum optimization~\cite{Cerezo_2022,Abbas_2024,Fontana_2024}, we concentrate on whether the PQC architecture can express the complexities of the target integral's structure, therefore providing a suitable proposal PDF for IS. To that end, we assume ideal training conditions to carry out state preparation. By using a statevector simulator, we have exact access to all probabilities. Consequently, we can employ the KL divergence without encountering shot-induced trainability issues.

\begin{figure*}[t!]
\begin{center} 
\captionsetup{justification=Justified, singlelinecheck=off} 
\includegraphics[width=\textwidth]{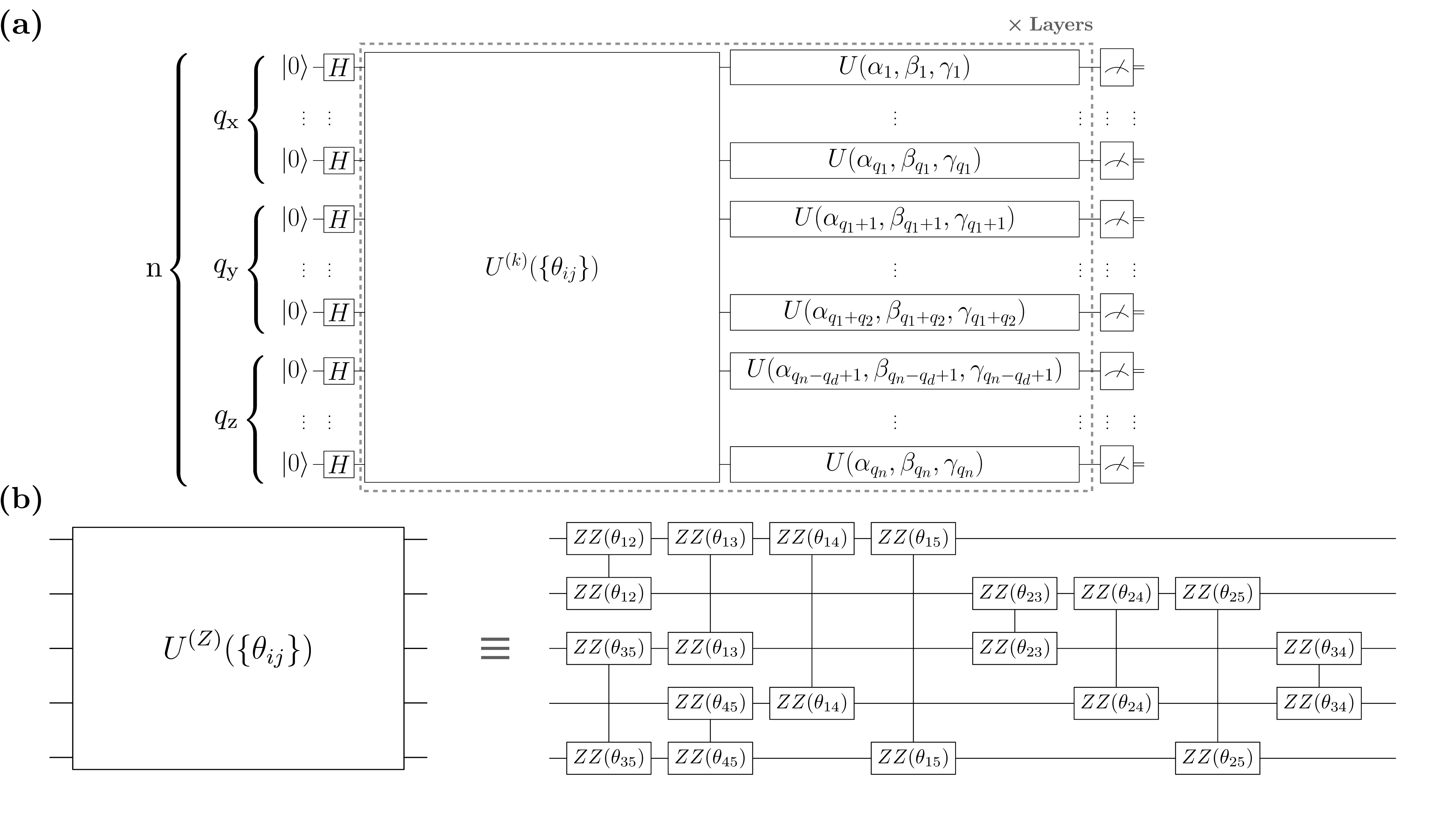}\caption{ {\bf Quantum Circuit architecture used for training.} {\bf (a)} General structure for a three-dimensional integrand with the layer definition consisting of a block of two-qubit gates, and a block of one-qubit rotations. The one-qubit block is denoted generically by $U$. For the \hyperref[sec:4.2.1]{One-loop pentagon Feynman integral in the Loop-Tree Duality} and \hyperref[sec:4.2.2]{Multi-peak Benchmark Integrals}, we use $U(\alpha,\beta,\gamma)=U3(\alpha,\beta,\gamma)$, while in \hyperref[sec:4.3]{Trainability for PQC State Preparation} we use $U(\alpha,\beta,\gamma)=R_Z(\alpha)R_X(\beta)R_Y(\gamma)$. {\bf (b)} Explicit $5$ qubit example of the $U^{(k)} (\{\theta_{ij}\})$ notation, with $k=Z$.
}
\label{fig:qc}
\end{center}
\end{figure*}

We proceed to introduce the architecture of the PQC used in the Results subsections \hyperref[sec:4.2.1]{One-loop pentagon Feynman integral in the Loop-Tree Duality} and \hyperref[sec:4.2.2]{Multi-peak Benchmark Integrals} . We employ an all-to-all connectivity approach that contains two-qubit gates and single qubit gates. Each two-qubit gate carries a tunable parameter, and the combination with the single qubit gates defines one layer. The form of the unitary operators are:
\begin{equation}
\begin{split}
    &U^{(k)}(\{ \theta_{ij} \}) = \prod_{i<j} e^{-i \theta_{ij} \sigma^{(k)}_i \sigma^{(k)}_j } ,\\
    & U3( \alpha_l,\beta_l,\gamma_l ) \propto e^{-i \beta_{l} \sigma^{(Z)}_l }e^{-i \alpha_{l} \sigma^{(Y)}_l }e^{-i \gamma_{l} \sigma^{(Z)}_l }~,
\end{split}
\end{equation}
where the first product runs over all distinct pairs of qubits and $k$ labels a distinct Pauli operator with $k \in \{X,Y,Z\}$. The total number of parameters per layer is $n(n-1)/2+3n\sim\mathcal{O}(n^2)$, where $n$ is the number of qubits. The architecture of the PQC is illustrated in Fig.~\ref{fig:qc}.

It is important to note that our main approach employs a highly-expressive multi-parameterized all-to-all connectivity Ansatz. For comparison, we also experimented with the popular Hardware Efficient Ansatz (HEA)~\cite{Schuld_2020,Leone_2024}, testing both its two-qubit CNOT gate version and its two-qubit parameterized gate version with its typical restricted nearest-neighbour connectivity. Based on our experiments, HEA performs significantly worse in this setting, not only for shallower PQCs but also when additional layers are added to match the number of parameters of our proposed Ansatz. Even then, its best cost function value would be substantially behind our proposed Ansatz.  

For optimization, we use COBYLA~\cite{Powell:1994xno,Powell_1998}  as well as the parameter-shift rule combined with Adam~\cite{kingma2017adammethodstochasticoptimization}, with all methods yielding similar outcomes in the proposal PDF quality. All experiments are performed using PennyLane's \texttt{lightning.qubit} noiseless state-vector simulator.

For completeness, we summarize here the optimization settings used in the numerical studies. For the pentagon integral,  the training of the PQC was carried out with COBYLA, and different Anz\"atze with different numbers of qubits per dimension. It is worth noting that, in the smallest configuration we tested, i.e. the 16 qubits distributed as $(8,4,4)$ across the three dimensions and $2 \times 10^4$ iterations, the Ansatz sequence $ U^{(Z)} \to U3$ with $144$ parameters reaches $D_{\rm KL} <0.5$ very fast and then saturates at $D_{\rm KL} \approx 0.27$. The results presented in Fig.~\ref{fig:pentagon_results0} are obtained using the Ansatz $U^{(Z)}  \to U3 \to U^{(X)}  \to U3$, which performed best given the constraints we imposed on the number of iterations. In this case, the number of parameters ranges from $336$ to $456$ for $16$ to $19$ qubits, respectively, and the number of iterations ranges from $2\times10^{4}$ to $3\times10^{4}$. The $D_{\rm KL}$ ranged from $0.09$ to $0.14$.

For the multi-peaked benchmark integrals we consider the same three-peak structure in spatial dimensions $d=2,3,4$, using the same discretization in all instances, which is five qubits per dimension. For the optimization, we employed $20 \times 10^3$ to $85 \times10^3$ COBYLA iterations for the PQC with $225$-$750$ parameters. The Ansatz chosen is the sequence $U^{(Z)} \to U3 \to U^{(Y)} \to U3 \to U^{(X)} \to U3$, which proved to be the best choice while considering the final KL divergence and the number of iterations. The resulting KL divergences fall in the range $0.09-0.21$.

%%%%%%%%%%%%%%%%%%%%%%%%%%%%%%%%%%%%%%%%%%%%%%%%%%%%%%%%%%%%%%%%%%%%%%%%%%%%%%%%%%%%% 
\subsection{Trainability Diagnostics} \label{sec:trainability_methods}

%%%%%%%%%%%%%%%%%%%%%%%%%%%%%%
%%%%%%%%%%%%%%%%%%%%%%%%%%%%%%%
 
We now describe the initialization strategy and diagnostics used in the trainability analysis, in Results subsection \hyperref[sec:4.3]{Trainability for PQC State Preparation}. The starting point of our initialization strategy is directly inherited from the standard practice in classical AIS as well as Normalizing Flows. In particular, classical approaches typically begin optimization from a uniform distribution and then adapt the PDF to capture the structure of the target density. In an $n$-qubit quantum computational setting, the natural analogue of a uniform PDF on the computational basis is the low-entanglement product state $|+\rangle^{\otimes n}$. Thus, in QAIS we initialize the PQC in this uniform regime and introduce randomness through small parameter perturbations around it, thereby providing a classically-inspired warm start from which the proposal PDF is adapted.
In particular, rather than choosing parameters randomly, we employ a restricted-angle initialization in which
\begin{equation}
\theta_i = \alpha\,\phi_i,\qquad \phi_i \sim \mathcal{N}(0,1),\qquad \alpha=0.01,
\end{equation}
so that $\theta_i$ are generated by a Gaussian distribution and are scaled down through the constant $\alpha$, keeping the parameters close to $0$. This initializes each parametrized gate close to the identity and keeps the PQC far from strongly entangled initial states. Restricted small-angle initialization approaches are discussed in the barren-plateau literature as a practical mitigation strategy~\cite{Larocca_2025,zhang2025escapingbarrenplateaugaussian,Wang_2024}. 
To demonstrate this approach, we minimize the KL divergence $D_{\text{KL}} (\boldsymbol{\theta})\equiv D_{\text{KL}}(P \| Q_{\boldsymbol{\theta}})$, as discussed in the Methods subsections \hyperref[sec:sec3.4]{Structure and Optimization of the Parametrized Quantum Circuit}. Two initialization scenarios are examined, one where the parameters are randomly initialized uniformly in their domain and the other in which they are initialized using restricted angles. Gradients are computed via the parameter-shift rule~\cite{Schuld_2019,Wierichs_2022}, through \texttt{Pennylane}'s predefined function. The optimizer used is Adam with learning rate $0.05$. The number of iterations is set through a hard cutoff at $100$. A single layer of the PQC used is the very expressive, heavily parameterized sequence $U^{(Z)} \to R_Z \to R_X \to R_Y \to U^{(Y)} \to R_Z \to R_X \to R_Y  \to U^{(X)} \to R_Z \to R_X \to R_Y $, with $R_{i}$ the standard single qubit rotation operations applied in each qubit with an independent parameter. We repeat this layer $L$ times.
To monitor training, we compute the gradient vector  at each optimization step. The simple parameter-wise average $\frac{1}{N_P} \sum_i \frac{\partial D_{\mathrm{KL}}(\boldsymbol{\theta})}{\partial\theta_i} = \frac{1}{N_P} \sum_i \nabla_i D_{\text{KL}} (\boldsymbol{\theta}) $, where $N_P$ is the number of trainable parameters, is not sufficiently informative because it is signed and typically suffers cancellations and tends to $0$ as $N_P$ grows, unless there is a strong bias toward one direction. To obtain a more meaningful measure of the gradient magnitude per iteration, we monitor the Root Mean Square (RMS) of the gradient components, 
\begin{equation}
\mathrm{RMS} \left( \boldsymbol{\nabla}   D_{\mathrm{KL}} (\boldsymbol{\theta} ) \right)=\sqrt{\frac{1}{N_P}\sum_{i=1}^{N_P}\left(\frac{\partial D_{\mathrm{KL}}(\boldsymbol{\theta})}{\partial\theta_i}\right)^2}.
\end{equation}
Based on this, we build and report the normalized gradient scale  $\mathrm{RMS} \left(\nabla D_{\mathrm{KL}}(\boldsymbol{\theta})\right)/D_{\mathrm{KL}}(\boldsymbol{\theta})$ as a dimensionless quantity for the effective update signal per unit loss value. This metric cannot, in general, distinguish vanishing gradient effects and stagnation to local minima. A direct barren plateau test would require the RMS on a single parameter across multiple random initializations, particularly on the first step of the optimization. However, it provides a practical indicator of how stable training is across different iterations, that would help in the transitioning to the shot-based case.

%%%%%%%%%%%%%%%%%%%%%%%%%%%%%%%%%%%%%%%%%%%%%%%%%%%%%%%%%%%%%%%%%%%%%%%%%%%%%%%%%%%%%%
\section*{Data availability} 
The trained parameters used to generate the results reported in this study are available at \url{https://github.com/Pyretz/qais}. Additional data supporting the findings of this study are available from the corresponding author upon request.

\section*{Code availability}
The code used in this work is open source and available at \url{https://github.com/Pyretz/qais}.

\section*{Author Contributions}
K.P. and G.R. conceived the project. K.P. designed the quantum algorithmic workflow and statistical framework with the Tiling algorithm and wrote the main codebase. J.M.L. implemented and optimised components of the main codebase and training pipelines. K.P. and J.M.L. executed the simulations. All authors contributed to the preparation and manuscript editing, the scientific discussions and approved the final manuscript. 

\section*{Competing interests}
The authors declare no competing interests.

%%%%%%%%%%%%%%%%%%%%%%%%%%%%%%%%%%%%%%%%%%%%%%%%%%%%%%%%%%%%%%%%%%%%%%%%%%%%%%%%%%%%%%

\begin{acknowledgments}
\section*{Acknowledgements}
This work is supported by the Spanish Government and ERDF/EU - Agencia Estatal de Investigaci\'on (MCIU/AEI/10.13039/501100011033), Grants No. PID2023-146220NB-I00, No. PID2020-114473GB-I00, No. EUR2025-164820, and No. CEX2023-001292-S; and Generalitat Valenciana, Grant No. ASFAE/2022/009 (Planes Complementarios de I+D+i, NextGenerationEU). KP is supported by Grants No. CEX2023-001292-S and No. ASFAE/2022/009. JML is supported by Generalitat Valenciana, Grant No. ACIF/2021/219. This work is also supported by the Ministry of Economic Affairs and Digital Transformation of the Spanish Government and NextGenerationEU through the Quantum Spain project, and by CSIC Interdisciplinary Thematic Platform (PTI+) on Quantum Technologies (PTI-QTEP+).
\end{acknowledgments}

%%%%%%%%%%%%%%%%%%%%%%%%%%%%%%%%%%%%%%%%%%%%%%%%%%%%%%%%%%%%
\section*{References}

\bibliography{references}

%%%%%%%%%%%%%%%%%%%%%%%%%%%%%%%%%%%%%%%%%%%%%%%%%%%%%%%%%%%%
\newpage

%%%%%%%%%%%%%%%%%%%%%%%%%%%%%%%%%%%%%%%%%%%%%%%%%%%%%%%%%%%%%%%%%%%%%%%%%%%%%%%%%%%%%%%%%%%%%%%%%%%%%%%%%%%%%%%%%%%%%%%%%%%%%%%%%%%%%%%%%%
\clearpage 
\onecolumngrid

\makeatletter
\setcounter{figure}{0}
\renewcommand \thesection{S\@arabic\c@section}
\renewcommand\thetable{S\@arabic\c@table}
\renewcommand \thefigure{S\@arabic\c@figure}
\makeatother

\section{Supplementary Information}

\subsection{Supplementary Methods: Tiling Algorithm}

The Tiling algorithm constructed to cover the $d$-dimensional space in Quantum Adaptive Importance Sampling~(QAIS) proceeds as follows. We consider a $d$-dimensional grid in which axis~$i$ is subdivided into $2^{q_i}$ uniform cells. The grid contains $2^{n}$ cells in total, with $n = \sum_{i=1}^{d} q_i$. Every grid cell is labeled by an $n$-bit string $s$, partitioned into the $d$ dimensions as:
\begin{equation}
s=s^{(n)}\to \bigl(s^{(q_d)}, s^{(q_{d-1})},\dots, s^{(q_1)}\bigr),
\qquad |s^{(q_i)}| = q_i , 
\end{equation}
where the full bitstring's sub-block $s^{(q_i)}$ encodes the coordinate on axis $i$. Throughout this work we adopt the Big Endian Encoding (BEE) \cite{1667115,Iaconis_2024} conventions, i.e. the left-most bit is most significant when converting $s$ to the integer
\begin{equation}
\text{int}(s^{(q_i)})\in \{0,\dots,2^{q_{i}}-1\}~.
\end{equation}
Under this encoding the mapping
\begin{equation}
  s \longmapsto (x_d,x_{d-1},\dots,x_1), 
  \qquad x_i\in\{0,\dots,2^{q_i}-1\}, 
\end{equation}
is one-to-one between bit strings and grid cell coordinates.

After performing $N$ measurements we obtain $M$ distinct outcome strings, sparsely distributed within the $2^n$ sized space:
\begin{equation}
\{(s_1,p(s_1)),\dots,(s_M,p(s_M))\},
\end{equation}
where each $p(s_k)$ is the empirical frequency of $s_k$. If the first and last possible states (namely the binary states $00 \dots 00$ and $11 \dots 11$) have not been observed, we manually insert them with zero probability. Sorting these strings in an ascending order costs $\mathcal{O}(M\log M)$, or $\mathcal{O}(n ~M\log M)$ if the comparison cost in $n$ bits is also included. For consecutive sorted strings $s_k<s_{k+1}$ we define the gap size:
\begin{equation}
\Delta_k = \text{int}(s_{k+1})-\text{int}(s_k)-1 .
\end{equation}

If $\Delta_k>0$ there are $\Delta_k$ unobserved cells strictly between $s_k$ and $s_{k+1}$. If $\Delta_k=0$ the two observed cells are adjacent. Then, we proceed with the greedy construction of the hyper-rectangles to cover the missing ranges. The main difficulty is to cover every gap with a few hyper-rectangles, depending only on $d$ and $n$ and the number of measured outcomes $M$, while not depending on the exponentially large pool of unmeasured states $2^{n}-M=\sum_k \Delta_k$. The main procedure (lines~31-49 of Algorithm~\ref{alg:tilgo}) expands greedily from the least-significant right-most dimension towards the most-significant left-most one, grouping whole blocks of cells whenever lower-order dimensions are already fully spanned.

Starting from the lower corner of a gap, the Tiling algorithm iteratively constructs a tile that never exceeds the boundary $2^{q_i}$ on any axis, never exceeds the remaining gap size $\Delta_k$, and expands the dimension~$i+1$ only after dimension $i$ has been filled from coordinate $0$ up to its boundary. In each dimension except the highest-order one, at most two separate expansions are made. The top dimension cannot have full expansions. Consequently, a single gap is covered by at most $2(d-1)+1$ hyper-rectangles. 

Once all gaps are tiled, we obtain an ordered list of measured points and hyper-rectangles that jointly cover the entire space. Collapsing the unmeasured tiles into the non-Important Region (as described in Methods subsection Debiasing Strategy using a Tiling Algorithm) is now straightforward.  Each empty tile is merged with the next measured state, the latter’s probability $p(s_{k+1})$ being assigned to the entire hyper-rectangle. If necessary, for the last interval, ( from $s_{M-1}$ to the binary state $11 \dots 11$ ), we merge it into the group containing $s_{M-1}$. This is the case when the final state is unobserved. The full pseudocode, including complexity hints for every operation, is given in Alg.~\ref{alg:tilgo}.

%%%%%%%%%%%%%%%%%%%%%%%%%%%%%%%%%%%%%%%%%%%%%%%%%%%%%%%%%%%%%%%%%%%%%%%%%%%%%%%%%%%%%%%%%%%%%%%%%%%%%%%%%%%%%%%%%%%%%%

\begin{algorithm}[H]
{\fontsize{9.5pt}{10pt}\selectfont
\caption{Tiling Algorithm for Multidimensional Spaces}
\label{alg:tilgo}
\begin{algorithmic}[1]
\State \textbf{Input:} 
    \begin{itemize}
        \item Grid dimensions: $grid\_dims \gets [2^{q_d}, \ldots ,2^{q_1}]$
        \item Dictionary of Measured States as Bitstrings and Probabilities : $S$
    \end{itemize}
\State \textbf{Output:} Dictionary, with tiles (in coordinate format), that cover the integration space.
\Statex  =========================================================
\State \textbf{Initialize Result Dictionary (Associated Array):} $F \gets \{ \}$
\State Sort the states: $S\gets \left\{ (s_1, p(s_1)), (s_2, p(s_2)), \dots, (s_M,p(s_M)) \right\}$ \Comment{$\mathcal{O}(n \; M \log M)$}
\For{$i \gets 1$ \textbf{to} $M$}\Comment{$\mathcal{O}(M)$} 
    \State $current\_coord \gets lin\_to\_coord(\text{int}(s_i))$ \Comment{To integer, to grid coordinate (BEE)}
    \State $end\_coord \gets  current\_coord$ 
       \State \textbf{Append} $((current\_coord,end\_coord), p(s_i))$ to $F$
    \If{$i<M$}
    \State $\Delta \gets \text{int}(s_{i+1}) - \text{int}(s_i) - 1$ \Comment{Cells to fill}
    \If{$\Delta > 0$}
        \State $current\_linear \gets  \text{int}(s_i) + 1$ 
        \State $tile\_intervals \gets \textsc{IntervalsGenerator}( grid\_dims,current\_linear, \Delta)$ \Comment{$\mathcal{O}(d^3 n^2 )$}
        \State \textbf{Append} $(tile\_intervals, 0)$ to $F$
    \EndIf
    \EndIf
\EndFor
\State \Return $F$
\Statex  =========================================================
\Procedure{IntervalsGenerator}{$grid\_dims, current\_linear, \Delta$}
    \State \textbf{Initialize:} $intervals \gets \emptyset$
    \While{$\Delta > 0$}  \Comment{\# of loops bounded by: $|intervals|\approx \mathcal{O}(d)$}
        \State $current\_coord \gets lin\_to\_coord(current\_linear)$ 
        \State $(tile\_shape, cells\_filled) \gets \textsc{GreedyMultidimExpand}(current\_coord,\ \Delta)$ \Comment{$\mathcal{O}(d^2 \, n^2 )$}
        \State $end\_coord \gets \left( current\_coord[i] + tile\_shape[i] - 1\right), \quad i=1,\dots,d $  
        \State \textbf{Append} $(current\_coord, end\_coord)$ to $intervals$
        \State $\Delta \gets \Delta - cells\_filled$   
        \State $current\_linear \gets current\_linear  + cells\_filled$   
    \EndWhile
    \State \Return $intervals$ \Comment{$|intervals| \leq 2*(d-1)+1$}
\EndProcedure 
\Statex  =====================================================
\Procedure{GreedyMultidimExpand}{$current\_coord, \Delta$}
    \State \textbf{Initialize:} $tile\_shape \gets [1,\ldots,1]$ \Comment{$|tile\_shape| = d$}
    \State $tile\_shape[d] \gets \min\{\Delta,  grid\_dims[d] - current\_coord[d]\}$   
    \For{$i \gets d-1$ \textbf{down to} $1$} \Comment{$\mathcal{O}(d)$}
        \If{$current\_coord[i+1] = 0$ \textbf{and} $tile\_shape[i+1] = grid\_dims[i+1]$}
            \State $comb\_vol \gets \prod_{j=i+1}^{d} grid\_dims[j]$  \Comment{$\mathcal{O}(d \, n^2 )$}
            \State $tile\_extend \gets \lfloor \Delta/comb\_vol \rfloor$ 
            \If{$tile\_extend > 0$}
                \State $tile\_shape[i] \gets \min\{grid\_dims[i] - current\_coord[i],  tile\_extend\}$  
            \Else
                \State $tile\_shape[i] \gets 1$
            \EndIf
        \Else
            \State $tile\_shape[i] \gets 1$
        \EndIf
    \EndFor
    \State $block\_vol \gets \prod_{i=1}^{d} tile\_shape[i]$  
    \State \Return $(tile\_shape, block\_vol)$
\EndProcedure  
\Statex  =====================================================
\end{algorithmic}
}
\end{algorithm}

\end{document}